\UseRawInputEncoding
\documentclass[11pt]{article}
\usepackage{amsmath,amssymb,amsthm,amsxtra,overpic,bbm,bm,epsfig,ulem,color,multirow}
\usepackage{braket}

\begin{document}

\begin{center}
{\Large A split seesaw model with hidden neutrinoless double beta decay \\
but successful leptogenesis
}
\end{center}

\vspace{0.05cm}

\begin{center}
{\bf Zhen-hua Zhao\footnote{zhaozhenhua@lnnu.edu.cn}} \\
{ $^1$ Department of Physics, Liaoning Normal University, Dalian 116029, China \\
$^2$ Center for Theoretical and Experimental High Energy Physics, Liaoning Normal University, Dalian 116029, China }
\end{center}

\vspace{0.2cm}

\begin{abstract}
In a paper by Asaka, Ishida and Tanaka \cite{hiding}, they proposed a novel possibility (which will be referred to as the AIT ansatz) that, in spite of the Majorana nature of neutrinos, the neutrinoless double beta ($0\nu \beta\beta$) decay can be hidden. In the original AIT model, the AIT ansatz is realized in the minimal seesaw model with two right-handed neutrinos which have a hierarchical mass structure: the lighter and heavier right-handed neutrinos are respectively much lighter and heavier than the typical Fermi-momentum scale of nuclei.
However, the original AIT model does not accommodate a successful leptogenesis. For this problem, in this paper we study a split seesaw model with one lighter right-handed neutrino but two heavier right-handed neutrinos which can realize the AIT ansatz and accommodate a successful leptogenesis simultaneously. We first give the condition on the neutrino Yukawa couplings for realizing the AIT ansatz, discuss its realization by employing an Abelian flavor symmetry, and study its implications for the mixing of the lighter right-handed neutrino with three left-handed neutrinos. We then successively study the implications for leptogenesis of the interesting scenarios where $M^{}_{\rm D}$ is a triangular matrix (which has maximally-restricted texture zeros, in line with the simplicity principle) or respects the $\mu$-$\tau$ reflection symmetry (which is well motivated by the experimental results), on top of the AIT ansatz.
\end{abstract}

\newpage

\section{Introduction}

As we know, the phenomena of neutrino oscillations show that neutrinos are massive and the lepton flavors are mixed \cite{xing}. The type-I seesaw mechanism \cite{seesaw} where the Standard Model (SM) is extended with some (at least two) right-handed neutrinos $N^{}_I$ (for $I=1, 2, 3$) is one of the most attractive way to generate the observed tiny neutrino masses. These newly introduced particles not only can have the usual Yukawa couplings with the left-handed neutrinos $\nu^{}_\alpha$ (for $\alpha = e, \mu, \tau$) and the Higgs field $H$ in the form of $(Y^{}_{\nu})^{}_{\alpha I} \overline \nu^{}_\alpha H N^{}_I $ with $(Y^{}_{\nu})^{}_{\alpha I}$ being the neutrino Yukawa couplings, which lead to the Dirac neutrino mass terms $(M^{}_{\rm D})^{}_{\alpha I}= (Y^{}_{\nu})^{}_{\alpha I} v $ with $ v = 174$ GeV being the vacuum expectation value of $H$, but also can have their own Majorana masses. In this paper, without loss of generality, we will work in the basis where the right-handed neutrino mass matrix is diagonal $M^{}_{\rm R} = {\rm diag}(M^{}_1, M^{}_2, M^{}_3)$ with $M^{}_I$ being the mass of $N^{}_I$. After integrating the heavy right-handed neutrinos out, one will obtain an effective Majorana mass matrix $M^{}_\nu \simeq - M^{}_{\rm D} M^{-1}_{\rm R} M^{T}_{\rm D}$ [whose elements will be denoted as $(M^{}_{\nu})^{}_{\alpha \beta}$] for the light neutrinos.

In the basis where the mass eigenstates of three charged leptons are identical with their flavor eigenstates, the lepton flavor mixing matrix $U$ will arise as the unitary matrix for diagonalizing $M^{}_\nu$:
\begin{eqnarray}
U^\dagger M^{}_\nu U^* =  D^{}_\nu =  {\rm diag}(m^{}_1, m^{}_2, m^{}_3) \;,
\label{1}
\end{eqnarray}
with $m^{}_i$ being three light neutrino masses. Under the standard parametrization, $U$ is expressed in terms of three lepton flavor mixing angles $\theta^{}_{ij}$ (for $ij=12, 13, 23$), the Dirac CP phase $\delta$ and two Majorana CP phases $\rho$ and $\sigma$:
\begin{eqnarray}
U  = \left( \begin{matrix}
c^{}_{12} c^{}_{13} & s^{}_{12} c^{}_{13} & s^{}_{13} e^{-{\rm i} \delta} \cr
-s^{}_{12} c^{}_{23} - c^{}_{12} s^{}_{23} s^{}_{13} e^{{\rm i} \delta}
& c^{}_{12} c^{}_{23} - s^{}_{12} s^{}_{23} s^{}_{13} e^{{\rm i} \delta}  & s^{}_{23} c^{}_{13} \cr
s^{}_{12} s^{}_{23} - c^{}_{12} c^{}_{23} s^{}_{13} e^{{\rm i} \delta}
& -c^{}_{12} s^{}_{23} - s^{}_{12} c^{}_{23} s^{}_{13} e^{{\rm i} \delta} & c^{}_{23}c^{}_{13}
\end{matrix} \right) \left( \begin{matrix}
e^{{\rm i}\rho} &  & \cr
& e^{{\rm i}\sigma}  & \cr
&  & 1
\end{matrix} \right) \;,
\label{2}
\end{eqnarray}
where the abbreviations $c^{}_{ij} = \cos \theta^{}_{ij}$ and $s^{}_{ij} = \sin \theta^{}_{ij}$ have been employed.

Neutrino oscillations are sensitive to three lepton flavor mixing angles, the Dirac CP phase, and the neutrino mass squared differences $\Delta m^2_{ij} \equiv m^2_i - m^2_j$. Thanks to the various neutrino oscillation experiments, these parameters have been measured with good degrees of accuracy, except that at present there is only a preliminary result for $\delta$. For convenience of the reader, the global-fit results for these parameters \cite{global,global2} are presented here in Table~\ref{tab1}. Note that the sign of $\Delta m^2_{31}$ remains unknown, thus allowing for two possible neutrino mass orderings: the normal ordering (NO) $m^{}_1 < m^{}_2 < m^{}_3$ and inverted ordering (IO) $m^{}_3 < m^{}_1 < m^{}_2$. In comparison, neutrino oscillations are insensitive to the absolute values of the neutrino masses and the Majorana CP phases. Their values can only be inferred from some non-oscillatory processes or cosmological observations. Unfortunately, so far there has not been any lower constraint on the lightest neutrino mass, nor any constraint on the Majorana CP phases.

As neutral particles, neutrinos are the unique candidate for Majorana fermions in the Standard Model. In fact, they would really be Majorana particles if their masses were indeed generated via the seesaw mechanism. In this sense, testing the (Majorana or Dirac) nature of neutrinos will not only tell us if there exist Majorana fermions in nature but also shed light on the origin of neutrino masses. At present, the most promising way to tackle this issue is to search for the neutrinoless double beta ($0\nu \beta\beta$) decay \cite{0nbb}. This decay can be mediated by massive Majorana neutrinos and its rate is controlled by the so-called effective Majorana neutrino mass $m^{}_{\beta \beta}$.
Here $m^{}_{\beta \beta}$ receives two contributions as
\begin{eqnarray}
m^{}_{\beta \beta} = m^\nu_{\rm \beta \beta} + m^N_{\rm \beta \beta} \;,
\label{3}
\end{eqnarray}
with
\begin{eqnarray}
m^\nu_{\beta \beta} = (M^{}_\nu)_{ee} =  - \sum^{}_{I} \frac{(M^{}_{\rm D})^2_{eI}}{M^{}_I}  \;,
\label{4}
\end{eqnarray}
and
\begin{eqnarray}
m^N_{\beta \beta} = \sum^{}_{I} \frac{\Lambda^{2}_{\beta}}{\Lambda^{2}_{\beta}+ M^2_I} \frac{(M^{}_{\rm D})^2_{eI}}{M^{}_I}  \;,
\label{5}
\end{eqnarray}
where $\Lambda^{}_{\beta} \sim {\cal{O}}(100)$ MeV denotes the typical scale of Fermi momentum of a nucleus. It is straightforward to see that when a certain right-handed neutrino is much heavier than $\Lambda^{}_{\beta}$, its contribution to $m^{}_{\beta \beta}$ via the $m^N_{\beta \beta}$ term is vanishingly small. On the other hand, when a certain right-handed neutrino is much lighter than $\Lambda^{}_{\beta}$, its contributions to $m^{}_{\beta \beta}$ via the $m^N_{\beta \beta}$ and $m^\nu_{\beta \beta}$ terms will cancel out each other \cite{cancel}.

In a paper by Asaka, Ishida and Tanaka \cite{hiding, hiding2}, they proposed a novel possibility (which will be referred to as the AIT ansatz) that, in spite of the Majorana nature of neutrinos, the $0\nu \beta\beta$ decay can be hidden. In the original AIT model, the AIT ansatz is realized in the minimal seesaw model with two right-handed neutrinos which have a hierarchical mass structure: the lighter ($N^{}_1$) and heavier ($N^{}_2$) right-handed neutrinos are respectively much lighter and heavier than $\Lambda^{}_{\beta}$ (i.e., $M^{}_1 \ll \Lambda^{}_{\beta} \ll M^{}_2$). In this case, the contributions of $N^{}_1$ to $m^{}_{\beta \beta}$ via the $m^N_{\beta \beta}$ and $m^\nu_{\beta \beta}$ terms will cancel out each other, while the contribution of $N^{}_2$ to $m^{}_{\beta \beta}$ via the $m^N_{\beta \beta}$ term is vanishingly small. Then, it is easy to see that, under the condition of $(M^{}_{\rm D})^{}_{e2} =0$, the contribution of $N^{}_2$ to $m^{}_{\beta \beta}$ via the $m^\nu_{\beta \beta}$ term will also vanish, rendering $m^{}_{\beta \beta}$ to be vanishingly small.

The AIT ansatz deserves attention in the sense that, in spite of the Majorana nature of neutrinos, the $0\nu \beta\beta$ decay might not take place. However, the original AIT model does not accommodate a successful leptogenesis \cite{leptogenesis, Lreview}, since the latter needs at least two right-handed neutrinos above the Davidson-Ibarra bound $\sim 10^{9}$ GeV in the case of the right-handed neutrino masses being not nearly degenerate \cite{DI}. For this problem, in this paper we study a split seesaw model with one lighter right-handed neutrino but two heavier right-handed neutrinos which can realize the AIT ansatz and accommodate a successful leptogenesis simultaneously. Here the split seesaw model is just the usual type-I seesaw model except that now the right-handed neutrino masses are strongly hierarchical (i.e., one is much smaller than $\Lambda^{}_{\beta}$ while the other two are above the Davidson-Ibarra bound and thus responsible for leptogenesis).

The rest part of this paper is organized as follows. In section~2, we give the condition on the neutrino Yukawa couplings for realizing the AIT ansatz, discuss its realization by employing an Abelian flavor symmetry, and study its implications for the mixings of the lighter right-handed neutrino with three left-handed neutrinos. In sections~3 and 4, we respectively study the implications for leptogenesis of the interesting scenarios of $M^{}_{\rm D}$ being a triangular matrix (which has maximally-restricted texture zeros, in line with the simplicity principle)  and respecting the $\mu$-$\tau$ reflection symmetry (which is well motivated by the experimental results), on top of the AIT ansatz. In section~5, we summarize our main results.

\begin{table}\centering
  \begin{footnotesize}
    \begin{tabular}{c|cc|cc}
     \hline\hline
      & \multicolumn{2}{c|}{Normal Ordering}
      & \multicolumn{2}{c}{Inverted Ordering }
      \\
      \cline{2-5}
      & bf $\pm 1\sigma$ & $3\sigma$ range
      & bf $\pm 1\sigma$ & $3\sigma$ range
      \\
      \cline{1-5}
      \rule{0pt}{4mm}\ignorespaces
       $\sin^2\theta^{}_{12}$
      & $0.318_{-0.016}^{+0.016}$ & $0.271 \to 0.370$
      & $0.318_{-0.016}^{+0.016}$ & $0.271 \to 0.370$
      \\[1mm]
       $\sin^2\theta^{}_{23}$
      & $0.566_{-0.022}^{+0.016}$ & $0.441 \to 0.609$
      & $0.566_{-0.023}^{+0.018}$ & $0.446 \to 0.609$
      \\[1mm]
       $\sin^2\theta^{}_{13}$
      & $0.02225_{-0.00078}^{+0.00055}$ & $0.02015 \to 0.02417$
      & $0.02250_{-0.00076}^{+0.00056}$ & $0.02039 \to 0.02441$
      \\[1mm]
       $\delta/\pi$
      & $1.20_{-0.14}^{+0.23}$ & $0.80 \to 2.00$
      & $1.54_{-0.13}^{+0.13}$ & $1.14 \to 1.90$
      \\[3mm]
       $\Delta m^2_{21}/(10^{-5}~{\rm eV}^2)$
      & $7.50_{-0.20}^{+0.22}$ & $6.94 \to 8.14$
      & $7.50_{-0.20}^{+0.22}$ & $6.94 \to 8.14$
      \\[3mm]
       $|\Delta m^2_{31}|/(10^{-3}~{\rm eV}^2)$
      & $2.56_{-0.04}^{+0.03}$ & $2.46 \to 2.65$
      & $2.46_{-0.03}^{+0.03}$ & $2.37 \to 2.55$
      \\[2mm]
      \hline\hline
    \end{tabular}
  \end{footnotesize}
  \caption{The best-fit values, 1$\sigma$ errors and 3$\sigma$ ranges of six neutrino
oscillation parameters extracted from a global analysis of the existing
neutrino oscillation data \cite{global}. }
\label{tab1}
\end{table}

\section{Condition on neutrino Yukawa couplings for hiding $0\nu \beta\beta$ decay}

In our framework, the mass spectrum of three right-handed neutrinos is taken to be $M^{}_1 \ll \Lambda^{}_{\beta} \ll M^{}_2 < M^{}_3$. In this scenario, for the following form of $M^{}_{\rm D}$ [i.e., $(M^{}_{\rm D})^{}_{e2} = (M^{}_{\rm D})^{}_{e3} =0$]
\begin{eqnarray}
M^{}_{\rm D}  = \left( \begin{matrix}
a^{}_1 \sqrt{M^{}_1} & 0 & 0 \cr
a^{}_2 \sqrt{M^{}_1} & b^{}_2 \sqrt{M^{}_2}  & c^{}_2 \sqrt{M^{}_3} \cr
a^{}_3 \sqrt{M^{}_1} & b^{}_3 \sqrt{M^{}_2} & c^{}_3 \sqrt{M^{}_3}
\end{matrix} \right)  \;,
\label{2.1}
\end{eqnarray}
$m^{}_{\beta \beta}$ would be vanishingly small, realizing the AIT ansatz. This point can be easily understood with the help of Eqs.~(\ref{3}-\ref{5}): for $N^{}_1$, due to $M^{}_1 \ll \Lambda^{}_{\beta}$, its contributions to $m^{}_{\beta \beta}$ via the $m^N_{\beta \beta}$ and $m^\nu_{\beta \beta}$ terms will cancel out each other. For $N^{}_2$ and $N^{}_3$, their contributions to $m^{}_{\beta \beta}$ via the $m^N_{\beta \beta}$ term are  vanishingly small due to $\Lambda^{}_{\beta} \ll M^{}_{2, 3}$, and those via the $m^\nu_{\beta \beta}$ term will also vanish due to $(M^{}_{\rm D})^{}_{e2} = (M^{}_{\rm D})^{}_{e3} =0$. For such a form of $M^{}_{\rm D}$, $N^{}_1$ couples to all the three lepton flavors, while $N^{}_2$ and $N^{}_3$ only couple to the $\mu$ and $\tau$ flavors. In the literature, the texture zeros of a mass matrix can be ascribed to some flavor symmetry \cite{zero} or simply motivated by the Occam's razor hypothesis. In the following we take the ${\rm Z}^{}_{12}$ symmetry as an example to illustrate how the texture zeros of $M^{}_{\rm D}$ in Eq.~(\ref{2.1}) can be realized from some flavor symmetry. The reader who is not interested in the flavor-symmetry model building can simply skip this part (which has no connection with the the other parts).

The ${\rm Z}^{}_{12}$ symmetry takes effect in a way as follows: first of all, each relevant field should transform in a certain way under the action of the symmetry operation. To be specific, we arrange that the left-handed lepton doublets $\overline L^{}_\alpha$, the right-handed charged leptons $\alpha$ and the right-handed neutrinos transform in the following ways under the ${\rm Z}^{}_{12}$ symmetry
\begin{eqnarray}
& & \overline L^{}_e \to \omega \overline L^{}_e \;, \hspace{1.25cm} e \to \omega e \;, \hspace{1.25cm} N^{}_1 \to \omega N^{}_1 \;, \nonumber \\
& & \overline L^{}_\mu \to \omega^3 \overline L^{}_\mu \;, \hspace{1cm} \mu \to \omega^2 \mu \;, \hspace{1cm} N^{}_2 \to \omega^2 N^{}_2 \;, \nonumber \\
& & \overline L^{}_\tau \to \omega^8 \overline L^{}_\tau \;, \hspace{1cm} \tau \to \omega^5 \tau \;, \hspace{1cm} N^{}_3 \to \omega^5 N^{}_3 \;,
\label{2.2}
\end{eqnarray}
with $\omega = {\rm exp}({\rm i}\pi/6)$. Let us take the $\overline L^{}_e$ field as an example to illustrate the meaning of these transformations: under the action of the symmetry operation, $\overline L^{}_e$ transforms to $\omega \overline L^{}_e$. Then, in order to preserve the ${\rm Z}^{}_{12}$ symmetry, in the Lagrangian only the terms that keep invariant under the action of the symmetry operation as a whole are allowed, while the other terms would be forbidden.

From Eq.~(\ref{2.2}), it is easy to see that the bilinears $\overline L^{}_\alpha \beta$ and $\overline L^{}_\alpha N^{}_I$, respectively responsible for the charged lepton mass terms $(M^{}_l)^{}_{\alpha \beta}$ and the Dirac neutrino mass terms $(M^{}_{\rm D})^{}_{\alpha I}$, transform according to the following matrix
\begin{eqnarray}
\left( \begin{matrix}
\omega^2 & \omega^3 & \omega^6 \cr
\omega^4 & \omega^5 & \omega^8 \cr
\omega^9 & \omega^{10} & \omega
\end{matrix} \right)  \;,
\label{2.3}
\end{eqnarray}
while the bilinears $\overline {N^c_I} N^{}_J$, responsible for the right-handed neutrino mass terms $(M^{}_{\rm R})^{}_{IJ}$, transform according to the following matrix
\begin{eqnarray}
\left( \begin{matrix}
\omega^2 & \omega^3 & \omega^6 \cr
\omega^3 & \omega^4 & \omega^7 \cr
\omega^6 & \omega^7 & \omega^{10}
\end{matrix} \right)  \;.
\label{2.4}
\end{eqnarray}
In order to render a certain element of $M^{}_{\rm D}$ [e.g., $(M^{}_{\rm D})^{}_{e1}$] non-zero, one may introduce some singlet that has an appropriate transformation property under ${\rm Z}^{}_{12}$ [e.g., $\phi^{}_{e1} \to \omega^{10} \phi^{}_{e1}$] and will acquire a non-zero vacuum expectation value. If the singlets $\phi^{}_{e2}$ and $\phi^{}_{e3}$ that have the transformation properties $\phi^{}_{e2} \to \omega^{9} \phi^{}_{e2}$ and $\phi^{}_{e3} \to \omega^{6} \phi^{}_{e2}$ under ${\rm Z}^{}_{12}$ are absent, then the vanishing of $(M^{}_{\rm D})^{}_{e2}$ and $(M^{}_{\rm D})^{}_{e3}$ will be protected by the ${\rm Z}^{}_{12}$ symmetry.
On the other hand, $M^{}_l$ (and similarly for $M^{}_{\rm R}$) will be of the diagonal form as desired if there only exist $\phi^{}_{ee}$, $\phi^{}_{\mu\mu}$ and $\phi^{}_{\tau\tau}$ that have the transformation properties $\phi^{}_{ee} \to \omega^{10} \phi^{}_{ee}$, $\phi^{}_{\mu\mu} \to \omega^{7} \phi^{}_{\mu\mu}$ and $\phi^{}_{\tau\tau} \to \omega^{11} \phi^{}_{\tau\tau}$ under ${\rm Z}^{}_{12}$. Besides, an auxiliary ${\rm Z}^{}_3$ symmetry is also needed in order to distinguish the singlets $\phi^{}_{\alpha \beta}$, $\phi^{}_{\alpha I}$ and $\phi^{}_{IJ}$ (responsible for the generations of the non-zero elements of $M^{}_l$, $M^{}_{\rm D}$ and $M^{}_{\rm R}$, respectively), under which the relevant fields have the following transformation properties
\begin{eqnarray}
& & \overline L^{}_\alpha \to \omega^\prime \overline L^{}_\alpha \;, \hspace{1.25cm} \alpha \to \omega^\prime \alpha \;, \hspace{1.25cm} N^{}_I \to N^{}_I \;, \nonumber \\
& & \phi^{}_{\alpha \beta} \to \omega^\prime \phi^{}_{\alpha \beta} \;, \hspace{1cm} \phi^{}_{\alpha I} \to \omega^{\prime 2} \phi^{}_{\alpha I} \;, \hspace{1cm} \phi^{}_{IJ} \to \phi^{}_{IJ} \;,
\label{2.5}
\end{eqnarray}
with $\omega^\prime = {\rm exp}(2{\rm i}\pi/3)$.

On the other hand, the smallness of $M^{}_1$ compared to $M^{}_2$ and $M^{}_3$ can be explained by means of the $U(1)$ Froggatt-Nielsen (FN) symmetry \cite{FN} \footnote{In Ref.~\cite{split}, the splitting among the right-handed neutrino masses is realized by invoking the extra dimensional theory.}. One may assign an FN number $n$ for $N^{}_1$ and introduce a singlet $\phi$ with the FN number $-1$. Then, the mass term of $N^{}_1$ will be subject to the suppression factor $(\langle \phi \rangle/\Lambda)^{2n}$ compared to those of $N^{}_2$ and $N^{}_3$, with $\Lambda$ being the cutoff scale of the FN mechanism. And the Yukawa couplings of $N^{}_1$ are subject to the suppression factor $(\langle \phi \rangle/\Lambda)^{n}$ compared to those of $N^{}_2$ and $N^{}_3$. On the whole, the contribution of $N^{}_1$ to the light neutrino masses is comparable to those of $N^{}_2$ and $N^{}_3$ \cite{sterile}. This is because in the seesaw formula the contributions of a right-handed neutrino to the light neutrino masses are inversely proportional to its mass but quadratically proportional to its Yukawa couplings.
Before proceeding, we would like to point out that although there is no strong connection between $N^{}_1$ and $N^{}_{2, 3}$ due to their huge mass hierarchies, the existence of $N^{}_1$ will indirectly affect the Yukawa couplings $N^{}_2$ and $N^{}_3$ can take. This is because all the three right-handed neutrinos can have significant contributions to $M^{}_\nu$ (i.e., the part of $M^{}_\nu$ that $N^{}_2$ and $N^{}_3$ can contribute depends on how much the contribution of $N^{}_1$ is) for which we have experimental information.

Then, we reconstruct the model parameters $a^{}_i$, $b^{}_i$ and $c^{}_i$ in Eq.~(\ref{2.1}) in terms of the low-energy neutrino observables (i.e., the lepton flavor mixing parameters and neutrino masses). This can be done by making a direct comparison between $M^{}_\nu$ obtained from the seesaw formula:
\begin{eqnarray}
M^{}_{\nu} \simeq - \left( \begin{matrix}
a^{2}_1 & a^{}_1 a^{}_2 & a^{}_1 a^{}_3 \cr
a^{}_1 a^{}_2 & a^{2}_2+b^{2}_2+c^{2}_2 & a^{}_2 a^{}_3 + b^{}_2 b^{}_3+ c^{}_2 c^{}_3 \cr
a^{}_1 a^{}_3 & a^{}_2 a^{}_3 + b^{}_2 b^{}_3+ c^{}_2 c^{}_3 & a^{2}_3+b^{2}_3+c^{2}_3
\end{matrix} \right)  \;,
\label{2.6}
\end{eqnarray}
and that obtained via the reconstruction relation $M^{}_\nu = U D^{}_\nu U^T$ [see Eq.~(\ref{1})]:
\begin{eqnarray}
(M^{}_\nu)^{}_{ee} & = & m^{}_1 e^{2{\rm i} \rho} c^2_{12} c^2_{13}+
m^{}_2 e^{2{\rm i} \sigma}s^2_{12} c^2_{13} + m^{}_3 s^{2}_{13} e^{-2{\rm i} \delta} \; ,
\nonumber \\
(M^{}_\nu)^{}_{\mu\mu} & = & m^{}_1 e^{2{\rm i} \rho} \left(s^{}_{12}
c^{}_{23} + c^{}_{12} s^{}_{23} s^{}_{13} e^{{\rm i} \delta} \right)^2
 + m^{}_2 e^{2{\rm i} \sigma}\left(c^{}_{12} c^{}_{23} - s^{}_{12} s^{}_{23} s^{}_{13} e^{{\rm i} \delta} \right)^2 + m^{}_3 c^2_{13} s^2_{23} \; ,
\nonumber \\
(M^{}_\nu)^{}_{\tau\tau} & = & m^{}_1 e^{2{\rm i} \rho}
\left(s^{}_{12} s^{}_{23} - c^{}_{12} c^{}_{23} s^{}_{13} e^{{\rm i} \delta} \right)^2
 + m^{}_2 e^{2{\rm i} \sigma}\left(c^{}_{12} s^{}_{23} + s^{}_{12} c^{}_{23} s^{}_{13} e^{{\rm i} \delta} \right)^2 + m^{}_3 c_{13}^2 c_{23}^2 \; ,
\nonumber \\
(M^{}_\nu)^{}_{e\mu} & = & -m^{}_1 e^{2{\rm i} \rho} c^{}_{12}
c^{}_{13} \left(s^{}_{12} c^{}_{23} + c^{}_{12}
s^{}_{23} s^{}_{13} e^{{\rm i} \delta} \right)
 + m^{}_2 e^{2{\rm i} \sigma}s^{}_{12} c^{}_{13} \left( c^{}_{12} c^{}_{23} -
s^{}_{12} s^{}_{23} s^{}_{13} e^{{\rm i} \delta} \right) \nonumber \\
&& + m^{}_3 c^{}_{13} s^{}_{23} s^{}_{13} e^{-{\rm i} \delta} \; ,
\nonumber \\
(M^{}_\nu)^{}_{e\tau} & = & m^{}_1 e^{2{\rm i} \rho} c^{}_{12} c^{}_{13}
\left(s^{}_{12} s^{}_{23} - c^{}_{12} c^{}_{23} s^{}_{13} e^{{\rm i} \delta} \right)
 -m^{}_2 e^{2{\rm i} \sigma}s^{}_{12} c^{}_{13} \left(c^{}_{12} s^{}_{23} +
s^{}_{12} c^{}_{23} s^{}_{13} e^{{\rm i} \delta} \right) \nonumber \\
&& + m^{}_3 c^{}_{13} c^{}_{23} s^{}_{13} e^{-{\rm i} \delta} \; ,
\nonumber \\
(M^{}_\nu)^{}_{\mu\tau} & = & -m^{}_1 e^{2{\rm i} \rho} \left(s^{}_{12}
s^{}_{23} - c^{}_{12} c^{}_{23} s^{}_{13} e^{{\rm i} \delta} \right) \left(
s^{}_{12} c^{}_{23}+ c^{}_{12} s^{} _{23} s^{}_{13} e^{{\rm i} \delta} \right)
\nonumber \\
&& -m^{}_2 e^{2{\rm i} \sigma} \left(c^{}_{12} s^{}_{23} +
s^{}_{12} c^{}_{23} s^{}_{13} e^{{\rm i} \delta} \right) \left(c^{}_{12}
c^{}_{23} - s^{}_{12} s^{}_{23} s^{}_{13} e^{{\rm i} \delta} \right)
 + m^{}_3 c^2_{13} c^{}_{23} s^{}_{23} \;.
\label{2.7}
\end{eqnarray}
From Eqs.~(\ref{2.6}, \ref{2.7}) we can see that the particular form of $M^{}_{\rm D}$ in Eq.~(\ref{2.1}) is not phenomenologically allowed to give a vanishingly small $(M^{}_\nu)^{}_{ee}$. This is because in the limit of $(M^{}_\nu)^{}_{ee} \to 0$, one has $a^{}_1 \to 0$ [i.e., $(M^{}_{\rm D})^{}_{e1} \to 0$] which in turn leads to the unrealistic result that the electron flavor completely decouples from three right-handed neutrinos.
Since there are totally 7 model parameters but $M^{}_\nu$ only contains 6 independent elements, the former can not be completely reconstructed in terms of the latter and we will be left with one free parameter.
It is useful to note that $M^{}_\nu$ in Eq.~(\ref{2.6}) keeps invariant with respect to the following transformation
\begin{eqnarray}
\left( \begin{array}{c} b^\prime_2 \cr c^\prime_2 \end{array} \right) = \left( \begin{array}{cc} \cos z & -\sin z \cr \sin z & \cos z \end{array} \right) \left( \begin{array}{c} b^{}_2 \cr c^{}_2 \end{array} \right) \;, \hspace{1cm}
\left( \begin{array}{c} b^\prime_3 \cr c^\prime_3 \end{array} \right) = \left( \begin{array}{cc} \cos z & -\sin z \cr \sin z & \cos z \end{array} \right) \left( \begin{array}{c} b^{}_3 \cr c^{}_3 \end{array} \right) \;,
\label{2.8}
\end{eqnarray}
with $z$ being a complex parameter.
With the help of this observation, the reconstruction result can be obtained in a way as follows: we first derive it in the particular scenario of $c^{}_2 =0$ (where $M^{}_{\rm D}$ is a triangular matrix as will be studied in the next section). A direct calculation gives
\begin{eqnarray}
&& a^{}_1 = {\rm i} \eta^{}_{a^{}_1} \sqrt{(M^{}_\nu)^{}_{ee}} \;, \hspace{1cm} a^{}_2 = {\rm i}  \eta^{}_{a^{}_1} \frac{(M^{}_\nu)^{}_{e\mu}}{\sqrt{(M^{}_\nu)^{}_{ee}}} \;, \hspace{1cm} a^{}_3 = {\rm i} \eta^{}_{a^{}_1} \frac{(M^{}_\nu)^{}_{e\tau}}{\sqrt{(M^{}_\nu)^{}_{ee}}} \;, \nonumber \\
&& b^{}_2 = \eta^{}_{b^{}_2} \sqrt{ \frac{ (M^{}_\nu)^{2}_{e\mu} - (M^{}_\nu)^{}_{ee} (M^{}_\nu)^{}_{\mu\mu} }{(M^{}_\nu)^{}_{ee}}  } \;, \nonumber \\
&& b^{}_3 = \eta^{}_{b^{}_2} \frac{(M^{}_\nu)^{}_{e\mu} (M^{}_\nu)^{}_{e\tau} - (M^{}_\nu)^{}_{ee} (M^{}_\nu)^{}_{\mu\tau} }{ \sqrt{ (M^{}_\nu)^{}_{ee} [ (M^{}_\nu)^{2}_{e\mu} - (M^{}_\nu)^{}_{ee} (M^{}_\nu)^{}_{\mu\mu}  ] } } \;, \nonumber \\
&& c^{}_3 = \eta^{}_{c^{}_3} \sqrt{ \frac{ (M^{}_\nu)^{2}_{e\tau} - (M^{}_\nu)^{}_{ee} (M^{}_\nu)^{}_{\tau\tau} }{(M^{}_\nu)^{}_{ee}} - \frac{ [(M^{}_\nu)^{}_{e\mu} (M^{}_\nu)^{}_{e\tau} - (M^{}_\nu)^{}_{ee} (M^{}_\nu)^{}_{\mu\tau} ]^2 }{  (M^{}_\nu)^{}_{ee} [ (M^{}_\nu)^{2}_{e\mu} - (M^{}_\nu)^{}_{ee} (M^{}_\nu)^{}_{\mu\mu}  ]  } } \;,
\label{2.9}
\end{eqnarray}
with $\eta^{}_{a^{}_1}, \eta^{}_{b^{}_2}, \eta^{}_{c^{}_3} = \pm 1$. Then, the reconstruction result in the generic scenario of $c^{}_2 \neq 0$ can be expressed as
\begin{eqnarray}
b^{}_2 & = & \eta^{}_{b^{}_2} \sqrt{ \frac{ (M^{}_\nu)^{2}_{e\mu} - (M^{}_\nu)^{}_{ee} (M^{}_\nu)^{}_{\mu\mu} }{(M^{}_\nu)^{}_{ee}}  } \cos z \;, \nonumber \\
c^{}_2 & = & \eta^{}_{b^{}_2} \sqrt{ \frac{ (M^{}_\nu)^{2}_{e\mu} - (M^{}_\nu)^{}_{ee} (M^{}_\nu)^{}_{\mu\mu} }{(M^{}_\nu)^{}_{ee}}  } \sin z  \;, \nonumber \\
b^{}_3 & = & \eta^{}_{b^{}_2} \frac{(M^{}_\nu)^{}_{e\mu} (M^{}_\nu)^{}_{e\tau} - (M^{}_\nu)^{}_{ee} (M^{}_\nu)^{}_{\mu\tau} }{ \sqrt{ (M^{}_\nu)^{}_{ee} [ (M^{}_\nu)^{2}_{e\mu} - (M^{}_\nu)^{}_{ee} (M^{}_\nu)^{}_{\mu\mu}  ] } } \cos z  \nonumber \\
&& - \eta^{}_{c^{}_3} \sqrt{ \frac{ (M^{}_\nu)^{2}_{e\tau} - (M^{}_\nu)^{}_{ee} (M^{}_\nu)^{}_{\tau\tau} }{(M^{}_\nu)^{}_{ee}} - \frac{ [(M^{}_\nu)^{}_{e\mu} (M^{}_\nu)^{}_{e\tau} - (M^{}_\nu)^{}_{ee} (M^{}_\nu)^{}_{\mu\tau} ]^2 }{  (M^{}_\nu)^{}_{ee} [ (M^{}_\nu)^{2}_{e\mu} - (M^{}_\nu)^{}_{ee} (M^{}_\nu)^{}_{\mu\mu}  ]  } } \sin z \;, \nonumber \\
c^{}_3 & = &  \eta^{}_{b^{}_2} \frac{(M^{}_\nu)^{}_{e\mu} (M^{}_\nu)^{}_{e\tau} - (M^{}_\nu)^{}_{ee} (M^{}_\nu)^{}_{\mu\tau} }{ \sqrt{ (M^{}_\nu)^{}_{ee} [ (M^{}_\nu)^{2}_{e\mu} - (M^{}_\nu)^{}_{ee} (M^{}_\nu)^{}_{\mu\mu}  ] } } \sin z \nonumber \\
&& + \eta^{}_{c^{}_3} \sqrt{ \frac{ (M^{}_\nu)^{2}_{e\tau} - (M^{}_\nu)^{}_{ee} (M^{}_\nu)^{}_{\tau\tau} }{(M^{}_\nu)^{}_{ee}} - \frac{ [(M^{}_\nu)^{}_{e\mu} (M^{}_\nu)^{}_{e\tau} - (M^{}_\nu)^{}_{ee} (M^{}_\nu)^{}_{\mu\tau} ]^2 }{  (M^{}_\nu)^{}_{ee} [ (M^{}_\nu)^{2}_{e\mu} - (M^{}_\nu)^{}_{ee} (M^{}_\nu)^{}_{\mu\mu}  ]  } }  \cos z \;,
\label{2.10}
\end{eqnarray}
while the expressions of $a^{}_i$ are same as in Eq.~(\ref{2.9}).
It is obvious that $z$ acts as the above-mentioned remaining free parameter. Before proceeding, we make a comparison between our result for the reconstruction of $M^{}_{\rm D}$ (in terms of the low-energy neutrino observables and some additional parameters) and the well-known Casas-Ibarra parametrization \cite{CI}. It is known that, in the Casas-Ibarra parametrization, an $M^{}_{\rm D}$ of the generic form can be reconstructed in terms of the low-energy neutrino observables and three additional parameters. This can be easily understood from the following number counting: an $M^{}_{\rm D}$ of the generic form contains 9 independent elements, while $M^{}_\nu$ (which encodes the low-energy neutrino observables) only contains 6 independent elements, so one needs three additional parameters to reconstruct the former in terms of the latter. In comparison, an $M^{}_{\rm D}$ of the  particular form in Eq.~(\ref{2.1}) only contains 7 independent elements (as a result of its two elements being vanishing), so we just need one additional parameter to reconstruct it in terms of the low-energy neutrino observables.

\begin{figure*}[t]
\centering
\includegraphics[width=6.7in]{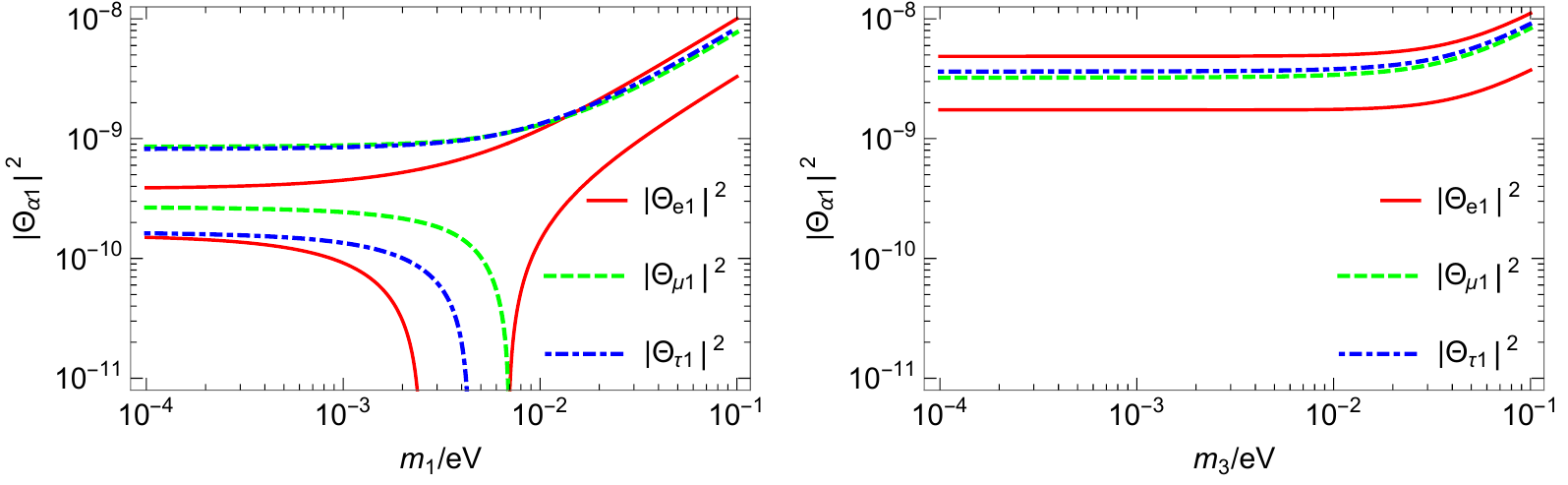}
\caption{ The maximally and minimally allowed values of $|\Theta^{}_{\alpha 1}|^2$ as functions of the lightest neutrino mass in the NO (left) and IO (right) cases. These results are obtained by taking $M^{}_1 = 10$ MeV and the best-fit values of the lepton flavor mixing angles and neutrino mass squared differences as typical inputs, and allowing the CP phases to vary in their whole ranges.}
\label{fig1}
\end{figure*}

Finally, we study the implications of $(Y^{}_\nu)^{}_{e2} = (Y^{}_\nu)^{}_{e3} =0$ for the mixings of $N^{}_1$ with three left-handed neutrinos. The magnitudes of these mixings directly determine the discovery prospects of $N^{}_1$ in relevant experiments \cite{rhn}. And the relative sizes of these mixings for three lepton flavors directly determine which flavor-specific channel will be the most promising one for the discovery of $N^{}_1$.
From the above reconstruction result, the mixing strengths $|\Theta^{}_{\alpha 1}|^2 \equiv |(M^{}_{\rm D})^{}_{\alpha1}|^2/M^{2}_1$ of $N^{}_1$ with $\nu^{}_{\alpha}$ are directly obtained as
\begin{eqnarray}
|\Theta^{}_{e 1}|^2 = \frac{|a^{}_1|^2}{M^{}_1} = \frac{|(M^{}_\nu)^{}_{ee}|}{M^{}_1} \;, \hspace{1cm}
|\Theta^{}_{\mu 1}|^2 = \frac{|a^{}_2|^2}{M^{}_1} = \frac{|(M^{}_\nu)^{}_{e\mu}|}{M^{}_1} \;, \hspace{1cm}
|\Theta^{}_{\tau 1}|^2 = \frac{|a^{}_3|^2}{M^{}_1} = \frac{|(M^{}_\nu)^{}_{e\tau}|}{M^{}_1} \;.
\label{2.11}
\end{eqnarray}
It is interesting that the magnitudes of $|\Theta^{}_{\alpha 1}|^2$ can be completely determined from the low-energy neutrino observables (plus $M^{}_1$). In Fig.~\ref{fig1}, we show the maximally and minimally allowed values of $|\Theta^{}_{\alpha 1}|^2$ as functions of the lightest neutrino mass in the NO and IO cases. These results are obtained by taking the best-fit values of the lepton flavor mixing angles and neutrino mass squared differences as typical inputs (same as below), and allowing the CP phases to vary in their whole ranges.  And $M^{}_1 = 10$ MeV has been taken as a benchmark value. Given that $|\Theta^{}_{\alpha 1}|^2$ are inversely proportional to $M^{}_1$, the results of the former for other values of the latter can be  obtained by rescaling the lines in Fig.~\ref{fig1} proportionally. From Fig.~\ref{fig1}, one can make the following observations. We first note that the behaviours of $|\Theta^{}_{\mu 1}|^2$ and $|\Theta^{}_{\tau 1}|^2$ are quite similar to each other. This fact is due to the approximate $\mu$-$\tau$ flavor symmetry in the neutrino sector \cite{mutau1, mutau2} which will be the subject of section~4.
In the NO case, $|\Theta^{}_{\mu 1}|^2$ and $|\Theta^{}_{\tau 1}|^2$ can be vanishingly small for sufficiently large $m^{}_1$. The maximally allowed values of $|\Theta^{}_{\mu 1}|^2$ and $|\Theta^{}_{\tau 1}|^2$ are about twice as large as (are approximately equal to) those of $|\Theta^{}_{e 1}|^2$ for sufficiently small (large) $m^{}_1$. In the IO case, $|\Theta^{}_{\mu 1}|^2$ and $|\Theta^{}_{\tau 1}|^2$ can be vanishingly small for the whole range of $m^{}_3$. The maximally allowed values of $|\Theta^{}_{\mu 1}|^2$ and $|\Theta^{}_{\tau 1}|^2$ are approximately equal to those of $|\Theta^{}_{e 1}|^2$ for the whole range of $m^{}_3$. And the maximally allowed values of $|\Theta^{}_{\alpha 1}|^2$ are much larger than in the NO case when the lightest neutrino mass is sufficiently small.

\section{Scenario of $M^{}_{\rm D}$ being a triangular matrix}

In this section, we consider the interesting scenario of $M^{}_{\rm D}$ being a triangular matrix obtained by taking $c^{}_2 =0$ in Eq.~(\ref{2.1}). Such a form of $M^{}_{\rm D}$ conforms to the simplicity principle in the sense that it has maximally-restricted texture zeros.
As discussed in section~2, for this scenario, the model parameters can be completely reconstructed from the low-energy neutrino observables as in Eq.~(\ref{2.9}). Then, we study the implications of this scenario for leptogenesis.

As we know, the type-I seesaw model also provides an attractive explanation (i.e., the leptogenesis mechanism) for the baryon-antibaryon asymmetry of the Universe \cite{planck}
\begin{eqnarray}
Y^{}_{\rm B} \equiv \frac{n^{}_{\rm B}-n^{}_{\rm \bar B}}{s} = (8.67 \pm 0.15) \times 10^{-11}  \;,
\label{3.1}
\end{eqnarray}
with $n^{}_{\rm B}$ ($n^{}_{\rm \bar B}$) being the baryon (antibaryon) number density and $s$ the entropy density. The leptogenesis mechanism works in a way as follows \cite{leptogenesis,Lreview}: a lepton-antilepton asymmetry $Y^{}_{\rm L} \equiv (n^{}_{\rm L}-n^{}_{\rm \bar L})/s$ is first generated from the CP-violating and out-of-equilibrium decays of the right-handed neutrinos and then partially transported to the baryon asymmetry: $Y^{}_{\rm B} \simeq - c Y^{}_{\rm L}$ with $c= 28/79$. In the unflavored regime which holds for $M^{}_I > 10^{12}$ GeV, the three lepton flavors are indistinguishable. In this regime, the final baryon asymmetry generated from the decays of $N^{}_I$ can be expressed as follows
\begin{eqnarray}
Y^{}_{I \rm B} = -c r \varepsilon^{}_I \kappa(\widetilde m^{}_I)  \;.
\label{3.2}
\end{eqnarray}
Here $r \simeq 3.9 \times 10^{-3}$ measures the ratio of the number density of $N^{}_I$ to the entropy density. $\varepsilon^{}_I$ is the total CP asymmetry for the decays of $N^{}_I$:
\begin{eqnarray}
\varepsilon^{}_{I } & = & \sum^{}_\alpha \varepsilon^{}_{I \alpha }= \sum^{}_\alpha \frac{  \left[ \Gamma(N^{}_I \to L^{}_\alpha + H) - \Gamma(N^{}_I \to \overline{L^{}_\alpha} + \overline{H}) \right]}{ \sum^{}_\alpha \left[\Gamma(N^{}_I \to L^{}_\alpha + H) + \Gamma(N^{}_I \to \overline{L^{}_\alpha} + \overline{H} ) \right] }  \nonumber \\
& = & \frac{1}{8\pi (M^\dagger_{\rm D}
M^{}_{\rm D})^{}_{II} v^2} \sum^{}_{J \neq I} {\rm Im}\left[ (M^\dagger_{\rm D} M^{}_{\rm D})^{2}_{IJ}\right] {\cal F} \left( \frac{M^2_J}{M^2_I} \right) \;,
\label{3.3}
\end{eqnarray}
while the flavored CP asymmetries are given by
\begin{eqnarray}
\varepsilon^{}_{I \alpha}
& = & \frac{1}{8\pi (M^\dagger_{\rm D}
M^{}_{\rm D})^{}_{II} v^2} \sum^{}_{J \neq I} \left\{ {\rm Im}\left[(M^*_{\rm D})^{}_{\alpha I} (M^{}_{\rm D})^{}_{\alpha J}
(M^\dagger_{\rm D} M^{}_{\rm D})^{}_{IJ}\right] {\cal F} \left( \frac{M^2_J}{M^2_I} \right) \right. \nonumber \\
&  &
+ \left. {\rm Im}\left[(M^*_{\rm D})^{}_{\alpha I} (M^{}_{\rm D})^{}_{\alpha J} (M^\dagger_{\rm D} M^{}_{\rm D})^*_{IJ}\right] {\cal G}  \left( \frac{M^2_J}{M^2_I} \right) \right\} \; ,
\label{3.4}
\end{eqnarray}
with ${\cal F}(x) = \sqrt{x} \{(2-x)/(1-x)+ (1+x) \ln [x/(1+x)] \}$ and ${\cal G}(x) = 1/(1-x)$.
And $\kappa(\widetilde m^{}_I) <1$ is the efficiency factor taking account of the effects of the processes (e.g., the inverse decays) that can washout the generated baryon asymmetry. Its value is determined by the so-called washout mass parameter
\begin{eqnarray}
\widetilde m^{}_I = \sum^{}_\alpha \widetilde m^{}_{I \alpha} = \sum^{}_\alpha  \frac{|(M^{}_{\rm D})^{}_{\alpha I}|^2}{M^{}_I} \; .
\label{3.42}
\end{eqnarray}
A detailed study shows that the dependence of the efficiency factor $\kappa(\widetilde m)$ on a certain washout mass parameter $\widetilde m$ can be well described by the following analytical fits \cite{giudice}
\begin{eqnarray}
\frac{1}{\kappa(\widetilde m)} \simeq \frac{3.3 \times 10^{-3} ~{\rm eV}}{\widetilde m} + \left( \frac{\widetilde m} {5.5 \times 10^{-4} ~{\rm eV}} \right)^{1.16} \;.
\label{3.5}
\end{eqnarray}
In the two-flavor regime which holds for $10^{12} > M^{}_I/{\rm GeV} > 10^9$, the $\tau$ flavor becomes distinguishable from the other two flavors which remain indistinguishable. In this regime, the final baryon asymmetry generated from the decays of $N^{}_I$ can be expressed as follows \cite{flavor}
\begin{eqnarray}
Y^{}_{I\rm B}
= - c r \left[ \varepsilon^{}_{I \tau} \kappa \left(\frac{390}{589} \widetilde m^{}_{I \tau} \right) + \varepsilon^{}_{I \gamma} \kappa \left(\frac{417}{589} \widetilde m^{}_{I \gamma} \right) \right] \;,
\label{3.6}
\end{eqnarray}
with $\varepsilon^{}_{I \gamma} = \varepsilon^{}_{I e} + \varepsilon^{}_{I \mu}$ and $\widetilde m^{}_{I \gamma} = \widetilde m^{}_{I e} + \widetilde m^{}_{I \mu}$. In the temperature range below $10^{9}$ GeV, all the three flavors are distinguishable and should be treated separately. As mentioned in the introduction section, the requirement for leptogenesis to be viable would place the Davidson-Ibarra lower bound $\sim 10^{9}$ GeV for the masses of the right-handed neutrinos responsible for leptogenesis \cite{DI}, hence we will just consider the unflavored and two-flavor regimes in the following discussions.

As mentioned above, in our framework it is $N^{}_2$ and $N^{}_3$ that will be responsible for leptogenesis. Depending on the relative sizes of $M^{}_{2, 3}$ and $10^{12}$ GeV (i.e., the boundary between the unflavored and two-flavor regimes), there are the following three possible leptogenesis scenarios. {\bf Scenario I}: for $M^{}_3 > M^{}_2 > 10^{12}$ GeV, the flavor effects have not come into play. After the $N^{}_3$-leptogenesis phase, the baryon asymmetry $Y^{0}_{3 \rm B}$ produced from the decays of $N^{}_3$ can be calculated according to Eq.~(\ref{3.2}) with
\begin{eqnarray}
\varepsilon^{}_{3} = \frac{M^{}_2 }{8\pi v^2 |c^{}_3|^2 }  {\rm Im}(b^{2}_3 c^{*2}_3) {\cal F} \left( \frac{M^2_2}{M^2_3} \right) \;, \hspace{1cm} \widetilde m^{}_3 = |c^{}_3|^2 \;.
\label{3.7}
\end{eqnarray}
Due to $M^{}_3 > M^{}_2$, $Y^{0}_{3 \rm B}$ is subject to the washout effects of $N^{}_2$. After the $N^{}_2$-leptogenesis phase, the surviving amount of $Y^{0}_{3 \rm B}$ is given by
\begin{eqnarray}
Y^{}_{3 \rm B} = \left[ p^{}_{32} \exp\left(-\frac{3\pi \widetilde m^{}_{2}}{8 m^{}_*}\right)  + 1- p^{}_{32} \right] Y^{0}_{3 \rm B} \;,
\label{3.8}
\end{eqnarray}
with $m^{}_* \simeq 1.1 \times 10^{-3}$ eV and
\begin{eqnarray}
p^{}_{32} =  \frac{ |(M^\dagger_{\rm D} M^{}_{\rm D})^{}_{32}|^2 }{ (M^\dagger_{\rm D} M^{}_{\rm D})^{}_{22} (M^\dagger_{\rm D} M^{}_{\rm D})^{}_{33} } = \frac{|b^{}_3|^2}{|b^{}_2|^2 + |b^{}_3|^2 }  \;.
\label{3.9}
\end{eqnarray}
Taking account of the baryon asymmetry $Y^{}_{2 \rm B}$ produced from the decays of $N^{}_2$, which can be calculated according to Eq.~(\ref{3.2}) with
\begin{eqnarray}
\varepsilon^{}_{2} = \frac{M^{}_3 }{8\pi  v^2 ( |b^{}_2|^2 + |b^{}_3|^2 )}  {\rm Im}(b^{*2}_3 c^{2}_3) {\cal F} \left( \frac{M^2_3}{M^2_2} \right) \;, \hspace{1cm} \widetilde m^{}_2 = |b^{}_2|^2 + |b^{}_3|^2 \;,
\label{3.10}
\end{eqnarray}
the final baryon asymmetry is given by $Y^{}_{ \rm B} =Y^{}_{2 \rm B} + Y^{}_{3 \rm B}$.

\begin{figure*}[t]
\centering
\includegraphics[width=6.7in]{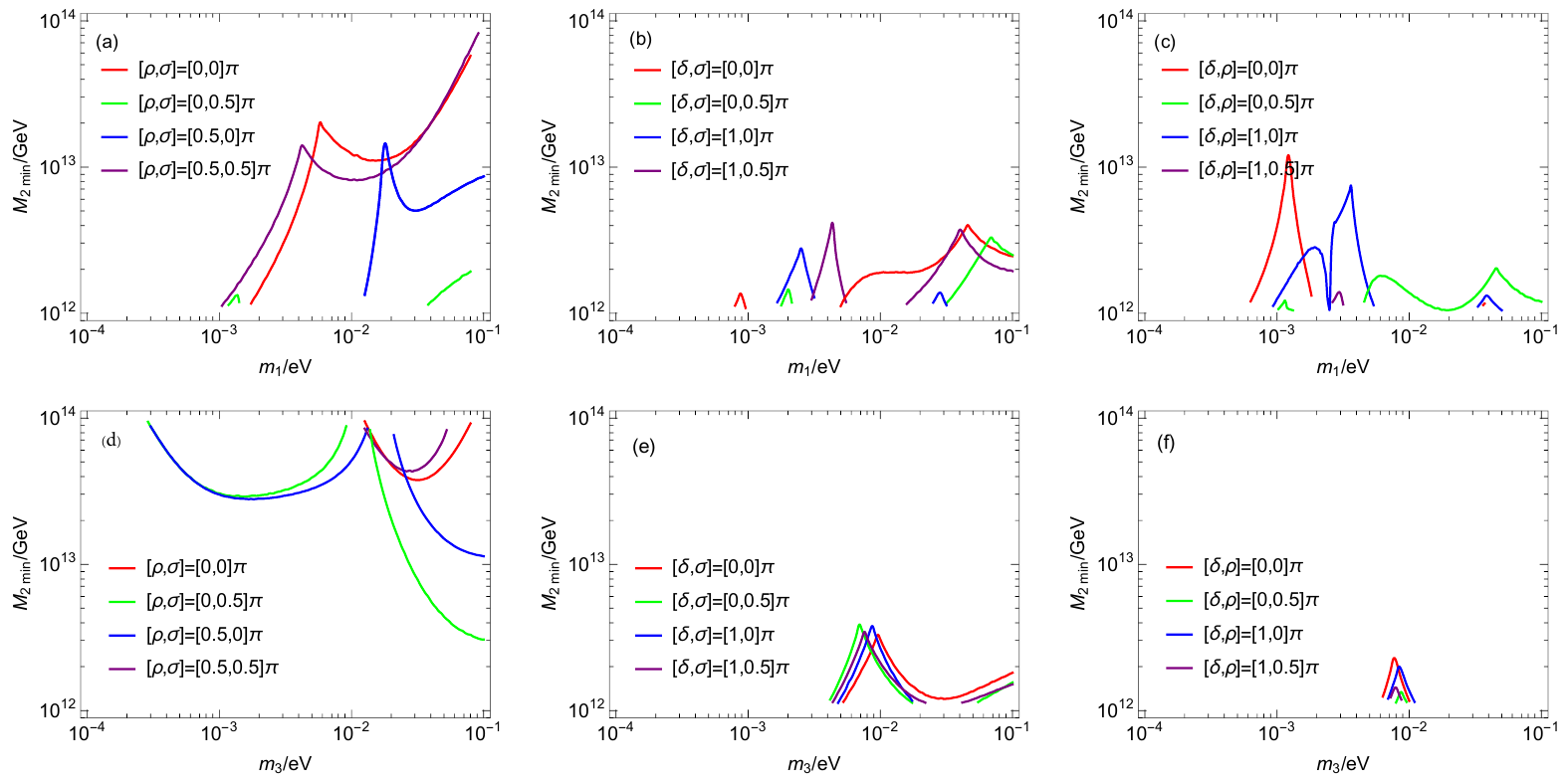}
\caption{ First column: in the scenario of $M^{}_{\rm D}$ being the triangluar matrix, for the various trivial-value combinations of $[\rho, \sigma]$, the lower bounds of $M^{}_2$ as functions of the lightest neutrino mass ($m^{}_1$ in the NO case, $m^{}_3$ in the IO case), imposed by requiring the {\bf Scenario I} leptogenesis to be successful. These results are obtained by allowing $\delta$ to vary in the whole range. Second (third) column: same as the first column, except that the roles of $\delta$ and $\rho$ ($\sigma$) are interchanged. }
\label{fig3.1}
\end{figure*}

\begin{figure*}[t]
\centering
\includegraphics[width=6.7in]{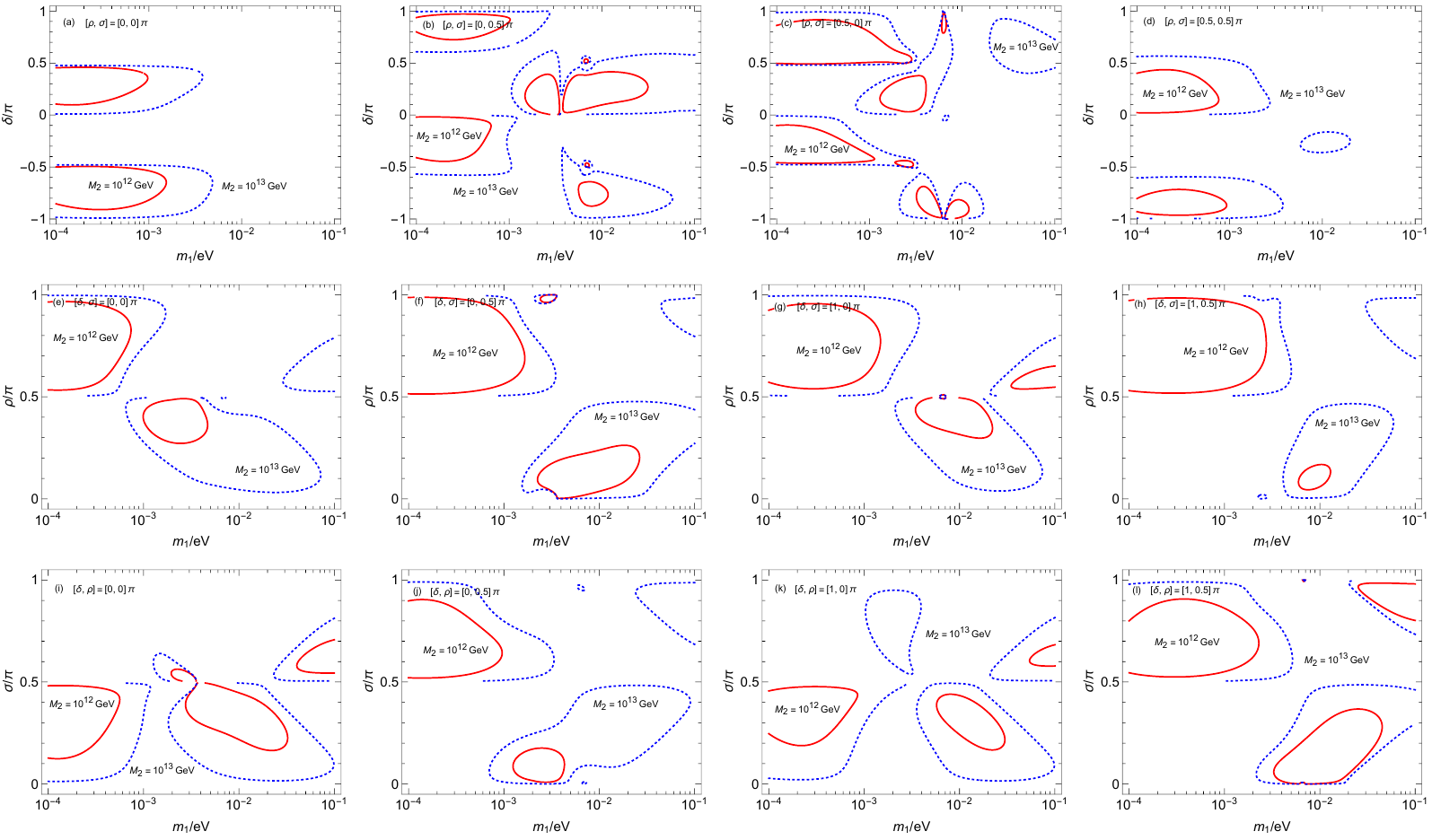}
\caption{ First row: in the scenario of $M^{}_{\rm D}$ being the triangluar matrix and NO case, for
the various trivial-value combinations of $[\rho, \sigma]$ and some benchmark values of $M^{}_2$ (see the labels for the plotted lines), the values of $\delta$ versus $m^{}_1$ for the {\bf Scenario I} leptogenesis to be successful. Second (third) row: same as the first row, except that the roles of $\delta$ and $\rho$ ($\sigma$) are interchanged. }
\label{fig3.2}
\end{figure*}

\begin{figure*}[t]
\centering
\includegraphics[width=6.7in]{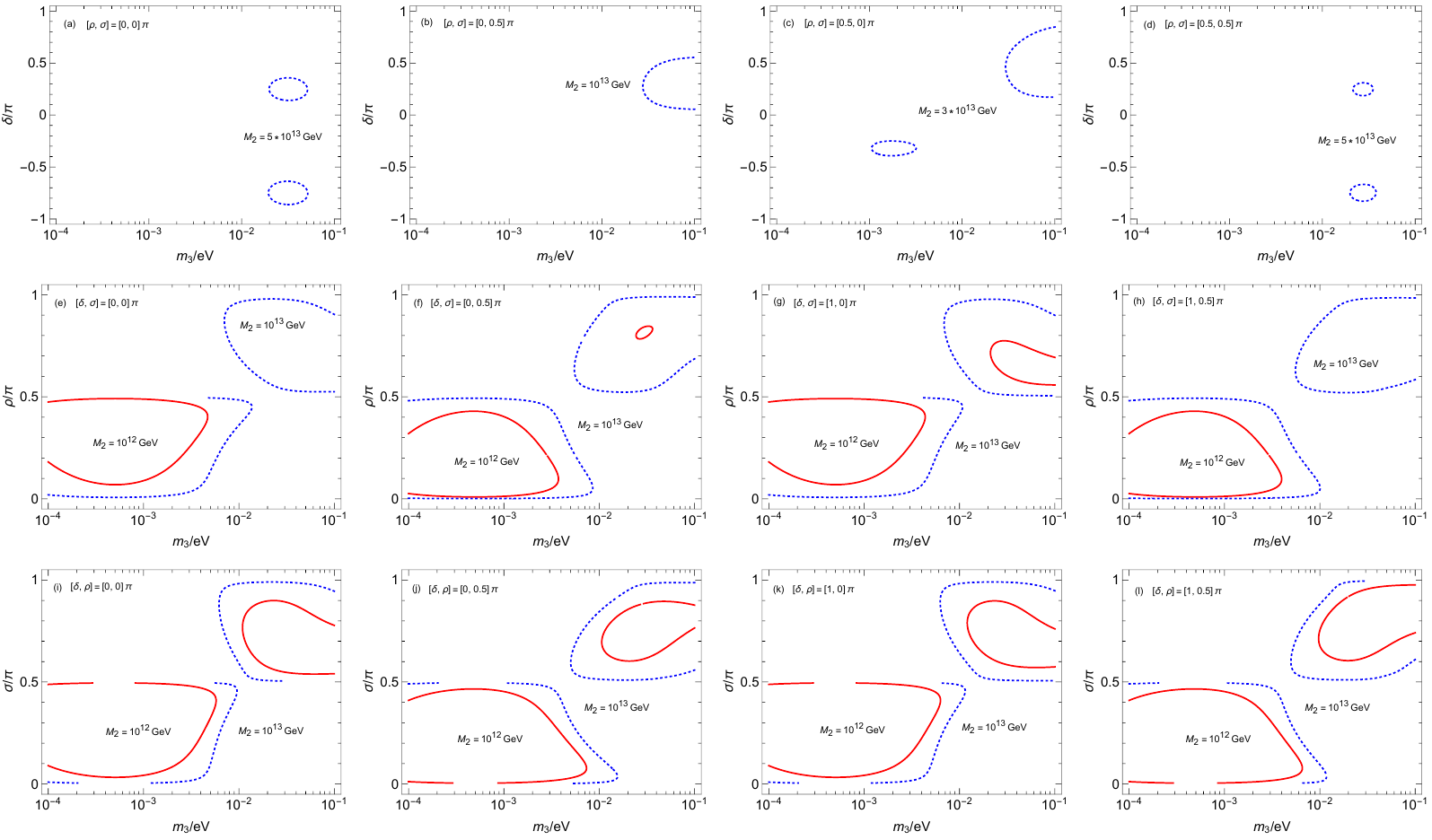}
\caption{ Same as Fig.~\ref{fig3.2}, except that the results are for the IO case. }
\label{fig3.3}
\end{figure*}

For the purpose of illustration, we only study the cases where only one of $\delta$, $\rho$ and $\sigma$ acts as the source for CP violation while the other two of them simply take trivial values (i.e., 0 or $\pi$ for $\delta$, 0 or $0.5\pi$ for $\rho$ and $\sigma$). In Fig.~\ref{fig3.1}, we plot the lower bounds of $M^{}_2$ as functions of the lightest neutrino mass ($m^{}_1$ in the NO case, $m^{}_3$ in the IO case), imposed by requiring the leptogenesis scenario under consideration to be successful.
These results are obtained by allowing one of $\delta$, $\rho$ and $\sigma$ to vary in the whole range, for the various trivial-value combinations of the other two of them. Note that when the right-handed neutrinos are heavier than $\sim 10^{14}$ GeV, their contributions to leptogenesis would be exponentially suppressed \cite{giudice}, so we have just shown the results for $M^{}_2 < 10^{14}$ GeV.
From these results one can make the following observations. In the NO case, when $\delta$ is the only source for CP violation, for $[\rho, \sigma]=[0, 0], [0.5, 0]\pi$ and $[0.5, 0.5]\pi$, there exist no constraints on $M^{}_2$ for small values of $m^{}_1$ but the lower bounds of $M^{}_2$ increase rapidly for large values of $m^{}_1$. For $[\rho, \sigma]=[0, 0.5]\pi$, there almost exist no constraints on $M^{}_2$ in the whole $m^{}_1$ range. When $\rho$ or $\sigma$ is the only source for CP violation, only in some small ranges of $m^{}_1$ there exist relatively small lower bounds for $M^{}_2$. These results indicate that leptogenesis is more difficult to be successful when $\delta$ is the only source for CP violation. This can be easily understood from that the effects of $\delta$ are always suppressed by $s^{}_{13}$.
In the IO case, the results are similar. When $\delta$ is the only source for CP violation, the ranges of $M^{}_2$ for leptogenesis to be successful are very small. But when $\rho$ or $\sigma$ is the only source for CP violation, there almost exist no constraints on $M^{}_2$. Furthermore, to see more clearly the constraints on the CP phases from leptogenesis, for some benchmark values of $M^{}_2$ (between the lower bounds of $M^{}_2$ obtained above and $10^{14}$ GeV), in Figs.~\ref{fig3.2} and \ref{fig3.3} (for the NO and IO cases, respectively) we plot the values of the CP phases versus the lightest neutrino mass for leptogenesis to be successful. We see that when $\delta$ is the only source for CP violation, the results for $[\rho, \sigma] = [0, 0]$ (or $[0, 0.5]\pi$) are similar to those for $[\rho, \sigma] = [0.5, 0.5]\pi$ (or $[0.5, 0]\pi$). When $\rho$ or $\sigma$ is the only source for CP violation, the results for $\delta =0$ are similar to those for $\delta =\pi$. This is also due to the suppression effect of $s^{}_{13}$ for the effects of $\delta$. Furthermore, in the IO case, the results for the case of $\rho$ being the only source of CP violation are similar to those for the case of $\sigma$ being the only source of CP violation. This is simply because one has $m^{}_1 \approx m^{}_2$ in the IO case.

\begin{figure*}[t]
\centering
\includegraphics[width=6.7in]{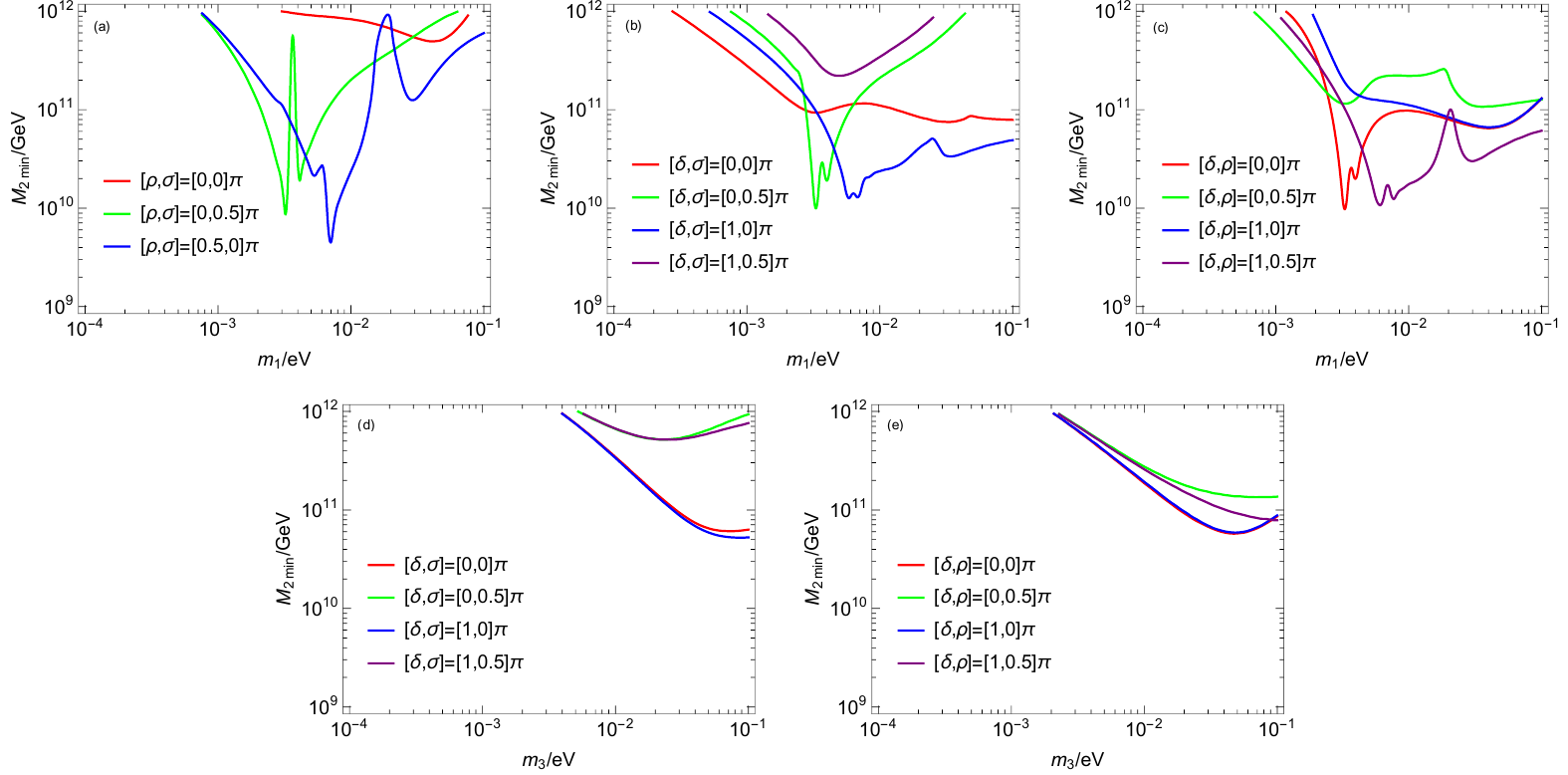}
\caption{ Same as Fig.~\ref{fig3.1}, except that the results are for the {\bf Scenario II} leptogenesis. }
\label{fig3.4}
\end{figure*}

\begin{figure*}[t]
\centering
\includegraphics[width=6.7in]{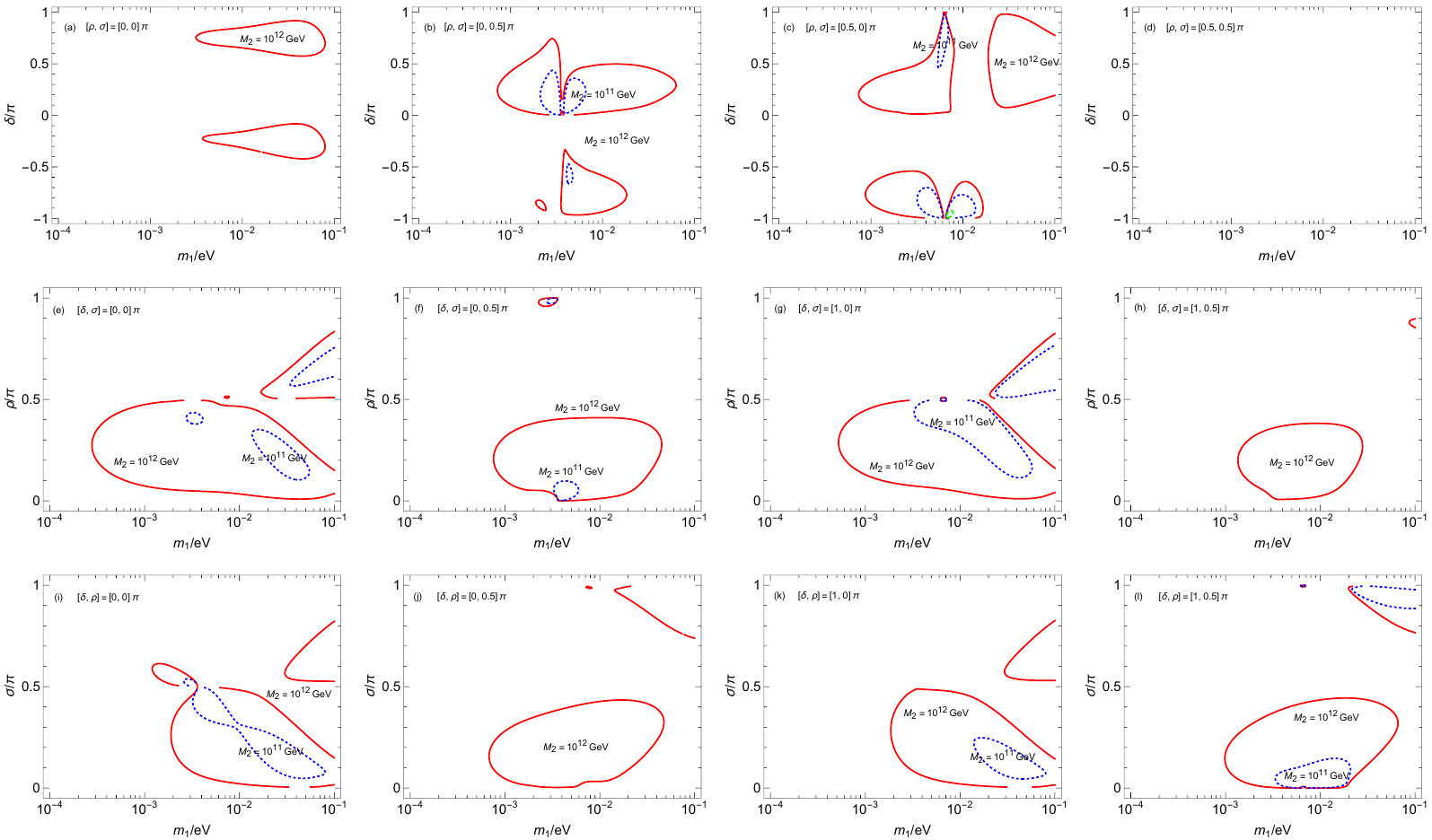}
\caption{ Same as Fig.~\ref{fig3.2}, except that the results are for the {\bf Scenario II} leptogenesis. }
\label{fig3.5}
\end{figure*}

\begin{figure*}[t]
\centering
\includegraphics[width=6.7in]{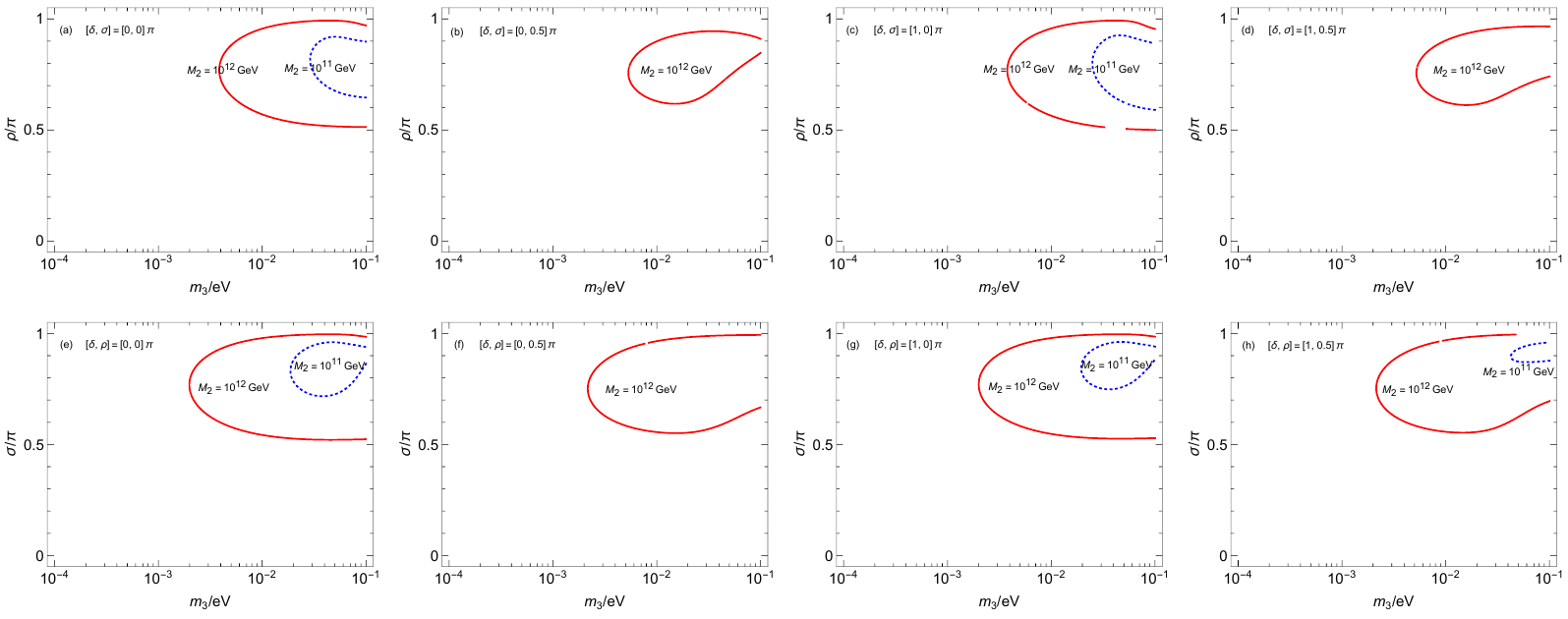}
\caption{ First row: in the scenario of $M^{}_{\rm D}$ being the triangluar matrix and IO case, for the
various trivial-value combinations of $[\delta, \sigma]$ and some benchmark values of $M^{}_2$ (see the labels for the plotted lines), the values of $\rho$ versus $m^{}_3$ for the {\bf Scenario II} leptogenesis to be successful. Second row: same as the first row, except that the roles of $\rho$ and $\sigma$ are interchanged. }
\label{fig3.6}
\end{figure*}

{\bf Scenario II}: for $M^{}_3 > 10^{12}$ GeV $>M^{}_2$, after the $N^{}_3$-leptogenesis phase, the baryon asymmetry $Y^{0}_{3 \rm B}$ produced from the decays of $N^{}_3$ can be calculated in a same way as in {\bf Scenario I}. After the $N^{}_2$-leptogenesis phase, the surviving amount of $Y^{0}_{3 \rm B}$ is given by
\begin{eqnarray}
Y^{}_{3\rm B} = Y^{0}_{3 \rm B} \exp\left(-\frac{3\pi \widetilde m^{}_{2\tau}}{8 m^{}_*}\right) \;,
\label{3.11}
\end{eqnarray}
with $\widetilde m^{}_{2\tau} = |b^{}_3|^2$.
And the baryon asymmetry produced from the decays of $N^{}_2$ can be calculated according to Eq.~(\ref{3.6}) with $\varepsilon^{}_{2 \tau} = \varepsilon^{}_{2}$ and $\varepsilon^{}_{2 \gamma} = 0$.
For this scenario, we have performed a similar analysis as for the former one. In Fig.~\ref{fig3.4}, we plot the lower bounds of $M^{}_2$ as functions of the lightest neutrino mass, imposed by requiring the leptogenesis scenario under consideration to be successful. From these results one can make the following observations. In the NO case, when $\delta$ is the only source for CP violation, leptogenesis has no (or very small) chance to be successful for $[\rho, \sigma]= [0.5, 0.5]\pi$ (or $[0, 0]$).
For $[\rho, \sigma]= [0, 0.5]\pi$ and $[0.5, 0]\pi$, leptogenesis has chance to be successful for $0.001 \lesssim m^{}_1/{\rm eV} \lesssim 0.1$ and $M^{}_2 \gtrsim 10^{10}$ GeV. When $\rho$ or $\sigma$ is the only source for CP violation, leptogenesis has chance to be successful for $m^{}_1 \gtrsim 0.001$ eV and $M^{}_2 \gtrsim 10^{10}$ GeV. In the IO case, when $\delta$ is the only source for CP violation, there exists no parameter space for leptogenesis to be successful at all. When $\rho$ or $\sigma$ is the only source for CP violation, leptogenesis has chance to be successful for $m^{}_1 \gtrsim 0.001$ eV and $M^{}_2 \gtrsim 10^{11}$ GeV. Similarly, for some benchmark values of $M^{}_2$ (between the lower bounds of $M^{}_2$ obtained above and $10^{12}$ GeV), in Figs.~\ref{fig3.5} and \ref{fig3.6} (for the NO and IO cases, respectively) we plot the values of the CP phases versus the lightest neutrino mass for leptogenesis to be successful. One can see that the observations made at the end of last paragraph basically still hold.

{\bf Scenario III}: for $ 10^{12}$ GeV $>M^{}_3 >M^{}_2$, after the $N^{}_3$-leptogenesis phase, the baryon asymmetry $Y^{0}_{3 \rm B}$ produced from the decays of $N^{}_3$ can be calculated according to Eq.~(\ref{3.6}) with
\begin{eqnarray}
\varepsilon^{}_{3 \tau} = \varepsilon^{}_{3} \;, \hspace{1cm} \varepsilon^{}_{3 \gamma} = 0 \;, \hspace{1cm}  \widetilde m^{}_{3 \tau} = |c^{}_3|^2 \;, \hspace{1cm} \widetilde m^{}_{3 \gamma} = 0 \;.
\label{3.12}
\end{eqnarray}
After the $N^{}_2$-leptogenesis phase, the surviving amount of $Y^{0}_{3 \rm B}$ and the baryon asymmetry produced from the decays of $N^{}_2$ can be calculated in a same way as in the former scenario. The numerical results for this scenario are very similar to those for the former one, so we will not explicitly show them.

\section{Scenario of $M^{}_{\rm D}$ respecting $\mu$-$\tau$ reflection symmetry}

Due to the closeness of the lepton flavor mixing angles to some special values (e.g., $\sin^2 \theta^{}_{23} \simeq 1/2$ and $\sin^2 \theta^{}_{12} \simeq 1/3$), it is believed by many people that there may exist some flavor symmetry in the lepton sector. And many flavor symmetry models have been proposed to explain the particular values of the lepton flavor mixing parameters \cite{FS}. One popular and attractive candidate of them is the $\mu$-$\tau$ reflection symmetry (which predicts $\theta^{}_{23} = \pi/4$ and $\delta = \pm \pi/2$) \cite{mu-tauR}. Hence in this section we consider the interesting scenario of $M^{}_{\rm D}$ in Eq.~(\ref{2.1}) also respecting the $\mu$-$\tau$ reflection symmetry and study the implications for leptogenesis.

The $\mu$-$\tau$ reflection symmetry is defined in a way as follows: the neutrino mass matrix should keep invariant with respect to the following transformations of three left-handed neutrino fields
\begin{eqnarray}
\nu^{}_{e} \leftrightarrow \nu^{c}_e \;, \hspace{1cm} \nu^{}_{\mu} \leftrightarrow \nu^{c}_{\tau} \;,
\hspace{1cm} \nu^{}_{\tau} \leftrightarrow \nu^{c}_{\mu} \;,
\label{4.1}
\end{eqnarray}
where the superscript ¡®$c$¡¯ denotes the charge conjugation of relevant neutrino fields.
Under this symmetry, the elements of $M^{}_\nu$ satisfy the following relations
\begin{eqnarray}
(M^{}_\nu)^{}_{e\mu} = (M^{}_\nu)^*_{e\tau} \;, \hspace{1cm} (M^{}_\nu)^{}_{\mu\mu} = (M^{}_\nu)^*_{\tau\tau}  \;, \hspace{1cm}
(M^{}_\nu)^{}_{ee} = (M^{}_\nu)^*_{ee} \;, \hspace{1cm} (M^{}_\nu)^{}_{\mu\tau} = (M^{}_\nu)^*_{\mu\tau} \;,
\label{4.2}
\end{eqnarray}
which lead to the following interesting predictions for the lepton flavor mixing parameters
\begin{eqnarray}
\theta^{}_{23} = \frac{\pi}{4}  \;, \hspace{1cm} \delta = \pm \frac{\pi}{2} \;,
\hspace{1cm} \rho, \sigma = 0 \ {\rm or} \ \frac{\pi}{2} \;.
\label{4.3}
\end{eqnarray}
Taking account of the results in Eq.~(\ref{4.3}), the expressions in Eq.~(\ref{2.7}) become
\begin{eqnarray}
&& (M^{}_\nu)^{}_{ee}  =  m^{}_1 \eta^{}_\rho c^2_{12} c^2_{13}+
m^{}_2  \eta^{}_\sigma s^2_{12} c^2_{13} - m^{}_3 s^{2}_{13}  \; ,
\nonumber \\
&& (M^{}_\nu)^{}_{e\mu} =  - \frac{1}{\sqrt 2}  m^{}_1 \eta^{}_\rho c^{}_{12} c^{}_{13} \left(s^{}_{12} \pm {\rm i} c^{}_{12} s^{}_{13} \right)
+ \frac{1}{\sqrt 2} m^{}_2 \eta^{}_\sigma s^{}_{12} c^{}_{13} \left( c^{}_{12} \mp {\rm i}
s^{}_{12} s^{}_{13} \right)  \mp \frac{1}{\sqrt 2} {\rm i} m^{}_3 c^{}_{13} s^{}_{13}  \; , \nonumber \\
&& (M^{}_\nu)^{}_{\mu\mu}  = \frac{1}{2}  m^{}_1 \eta^{}_\rho \left(s^{}_{12}
\pm {\rm i} c^{}_{12} s^{}_{13} \right)^2
+ \frac{1}{2}  m^{}_2 \eta^{}_\sigma \left(c^{}_{12} \mp {\rm i} s^{}_{12}
s^{}_{13} \right)^2 + \frac{1}{2}  m^{}_3 c^2_{13}  \; , \nonumber \\
&& (M^{}_\nu)^{}_{\mu\tau}  =  -\frac{1}{2}  m^{}_1 \eta^{}_\rho \left( s^{2}_{12} + c^{2}_{12} s^{2}_{13}  \right)
- \frac{1}{2} m_2^{} \eta^{}_\sigma \left(c^{2}_{12} + s^{2}_{12} s^{2}_{13} \right)
+ \frac{1}{2} m^{}_3 c^2_{13}   \; , \nonumber \\
&& (M^{}_\nu)^{}_{e\tau} = - (M^{}_\nu)^{*}_{e\mu},  \hspace{1cm} (M^{}_\nu)^{}_{\tau\tau} = (M^{}_\nu)^{*}_{\mu\mu}.
\label{4.4}
\end{eqnarray}
where $\eta^{}_\rho =1$ or $-1$ for $\rho=0$ or $\pi/2$ (and similarly for $\eta^{}_\sigma$), and the upper and lower signs of $\pm$ and $\mp$ respectively correspond to $\delta = \pi/2$ and $-\pi/2$.
Note that the sign difference for $(M^{}_\nu)^{}_{e\mu}/(M^{}_\nu)^{*}_{e\tau}$ in Eqs.~(\ref{4.2}, \ref{4.4}) is due to the unphysical phases.

Under the $\mu$-$\tau$ reflection symmetry, $M^{}_{\rm D}$ in Eq.~(\ref{2.1}) can be reexpressed in the following form
\begin{eqnarray}
M^{}_{\rm D}  = \left( \begin{matrix}
\sqrt{\eta^{}_1} a^{}_1 \sqrt{M^{}_1} & 0 & 0 \cr
\sqrt{\eta^{}_1} a^{}_2 \sqrt{M^{}_1} & \sqrt{\eta^{}_2} b^{}_2 \sqrt{M^{}_2}  & \sqrt{\eta^{}_3} c^{}_2 \sqrt{M^{}_3} \cr
\sqrt{\eta^{}_1} a^{*}_2 \sqrt{M^{}_1} & \sqrt{\eta^{}_2} b^{*}_2 \sqrt{M^{}_2} & \sqrt{\eta^{}_3} c^{*}_2 \sqrt{M^{}_3}
\end{matrix} \right)  \;,
\label{4.5}
\end{eqnarray}
with $a^{}_1$ being real now and $\eta^{}_I = \pm 1$. It is direct to verify that the resulting $M^{}_\nu$ from such an $M^{}_{\rm D}$
\begin{eqnarray}
M^{}_{\nu} \simeq - \left( \begin{matrix}
\eta^{}_1 a^{2}_1 & \eta^{}_1 a^{}_1 a^{}_2 & \eta^{}_1 a^{}_1 a^{*}_2 \cr
\eta^{}_1 a^{}_1 a^{}_2 & \eta^{}_1 a^{2}_2 + \eta^{}_2 b^{2}_2 + \eta^{}_3 c^{2}_2 & \eta^{}_1 |a^{}_2|^2  + \eta^{}_2 |b^{}_2|^2  + \eta^{}_3 |c^{}_2|^2  \cr
\eta^{}_1 a^{}_1 a^{*}_2 & \eta^{}_1 |a^{}_2|^2  + \eta^{}_2 |b^{}_2|^2  + \eta^{}_3 |c^{}_2|^2  & \eta^{}_1 a^{*2}_2 + \eta^{}_2 b^{*2}_2 + \eta^{}_3 c^{*2}_2
\end{matrix} \right)  \;,
\label{4.6}
\end{eqnarray}
does obey the relations in Eq.~(\ref{4.2}). As before, one can reconstruct the model parameters $a^{}_i$, $b^{}_i$ and $c^{}_i$ in Eq.~(\ref{4.5}) in terms of the low-energy neutrino observables by making a direct comparison between $M^{}_\nu$ in Eq.~(\ref{4.6}) and that described by Eq.~(\ref{4.4}). Note that there are totally 7 degrees of freedom among the model parameters but $M^{}_\nu$ only contains 6 degrees of freedom. Hence the model parameters can not be completely reconstructed in terms of the low-energy neutrino observables and we will be left with one degree of freedom.
Taking account of the effect of the unphysical phases as mentioned below Eq.~(\ref{4.4}), we obtain the following reconstruction result
\begin{eqnarray}
&& \sqrt{\eta^{}_1} a^{}_1 = {\rm i} \eta^{}_{a^{}_1} \sqrt{(M^{}_\nu)^{}_{ee}} \;, \hspace{1cm} \sqrt{\eta^{}_1} a^{}_2 = {\rm i}  \eta^{}_{a^{}_1} \frac{(M^{}_\nu)^{}_{e\mu}}{\sqrt{(M^{}_\nu)^{}_{ee}}}  \nonumber \\
&& \sqrt{\eta^{}_2} b^{}_2 =  \eta^{}_{b^{}_2} \sqrt{ \frac{ (M^{}_\nu)^{2}_{e\mu} - (M^{}_\nu)^{}_{ee} (M^{}_\nu)^{}_{\mu\mu} }{(M^{}_\nu)^{}_{ee}}  } \cos z \;, \nonumber \\
&& \sqrt{\eta^{}_3} c^{}_2 = \eta^{}_{b^{}_2} \sqrt{ \frac{ (M^{}_\nu)^{2}_{e\mu} - (M^{}_\nu)^{}_{ee} (M^{}_\nu)^{}_{\mu\mu} }{(M^{}_\nu)^{}_{ee}}  } \sin z  \;, \nonumber \\
&&  \eta^{}_2 |\cos^2 z| + \eta^{}_3 |\sin^2z| = \left| \frac{ (M^{}_\nu)^{}_{ee} }{  (M^{}_\nu)^{2}_{e\mu} - (M^{}_\nu)^{}_{ee} (M^{}_\nu)^{}_{\mu\mu} } \right| \left[ (M^{}_\nu)^{}_{\mu\tau} - \eta^{}_1 \left| \frac{(M^{}_\nu)^{2}_{e\mu}}{(M^{}_\nu)^{}_{ee}} \right| \right]  \;.
\label{4.7}
\end{eqnarray}
If $z$ is parameterized to be $x+{\rm i}y$ with $x$ and $y$ being real parameters, then one has
\begin{eqnarray}
\eta^{}_2 |\cos^2 z| + \eta^{}_3 |\sin^2z| = \pm \frac{1}{2} \left(e^{2y} + e^{-2y} \right) \hspace{0.3cm} {\rm or} \hspace{0.3cm} \pm \cos 2 x \;,
\label{4.8}
\end{eqnarray}
for $\eta^{}_2 = \eta^{}_3 = \pm 1$ or $\eta^{}_2 = - \eta^{}_3 = \pm 1$. These results show that we can always determine one of $x$ and $y$ from the low-energy neutrino observables, while the other one of them acts as the remaining degree of freedom.

Then, let us study the implications of the present scenario for leptogenesis. Substituting $M^{}_{\rm D}$ in Eq.~(\ref{4.5}) into the expressions of $\varepsilon^{}_{I }$ and $\varepsilon^{}_{I \alpha}$ in Eqs.~(\ref{3.3}, \ref{3.4}), we arrive at $\varepsilon^{}_{I } = \varepsilon^{}_{I e} =0$ and
\begin{eqnarray}
\varepsilon^{}_{2 \mu} = - \varepsilon^{}_{2 \tau}
= \frac{M^{}_3}{8 \pi v^2 |b^{}_2|^2}  {\rm Re}(b^{*}_2 c^{}_2)  {\rm Im}(b^*_2 c^{}_2 ) \left[ \eta^{}_2 \eta^{}_3 {\cal F} \left( \frac{M^2_3}{M^2_2} \right)
+  {\cal G}  \left( \frac{M^2_3}{M^2_2} \right) \right] \;, \nonumber \\
\varepsilon^{}_{3 \mu} = - \varepsilon^{}_{3 \tau}
= \frac{M^{}_2}{8 \pi v^2 |c^{}_2|^2}  {\rm Re}(b^{}_2 c^*_2)  {\rm Im}(b^{}_2 c^{*}_2 ) \left[ \eta^{}_2 \eta^{}_3 {\cal F} \left( \frac{M^2_2}{M^2_3} \right)
+  {\cal G}  \left( \frac{M^2_2}{M^2_3} \right) \right] \;.
\label{4.9}
\end{eqnarray}
As a result of $\varepsilon^{}_{I } =0$, in the unflavored regime, the baryon asymmetry produced from the decays of $N^{}_2$ and $N^{}_3$ is vanishing, making leptogenesis impossible. On the other hand, in the two-flavor regime, the baryon asymmetry produced from the decays of $N^{}_I$ (for $I=2, 3$) is given by
\begin{eqnarray}
Y^{}_{I\rm B}
= - c r \varepsilon^{}_{I \mu} \left[ \kappa \left(\frac{417}{589} \widetilde m^{}_{I \mu} \right) - \kappa \left(\frac{390}{589} \widetilde m^{}_{I \mu} \right)  \right] \;,
\label{4.10}
\end{eqnarray}
which is obtained from Eq.~(\ref{3.6}) by taking account of $\varepsilon^{}_{I e} = \widetilde m^{}_{I e} =0$, $\varepsilon^{}_{I \mu} = - \varepsilon^{}_{I \tau}$ and $\widetilde m^{}_{I \mu} = \widetilde m^{}_{I \tau}$. Thanks to the difference between the coefficients $417/589$ and $390/589$, a successful leptogenesis becomes possible.

\begin{figure*}[t]
\centering
\includegraphics[width=6.7in]{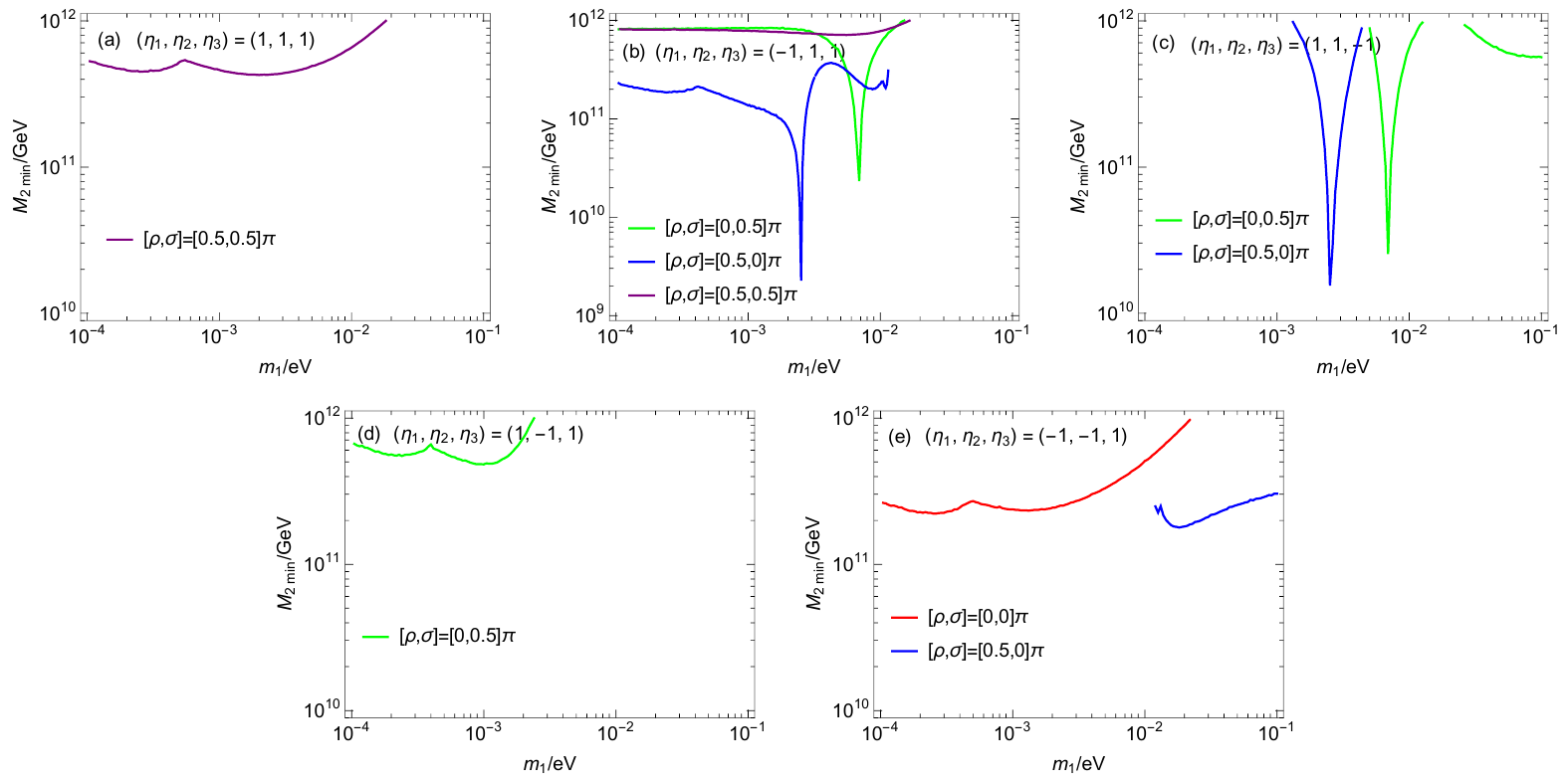}
\caption{ In the scenario of $M^{}_{\rm D}$ respecting the $\mu$-$\tau$ reflection symmetry, for the various combinations of $(\eta^{}_1, \eta^{}_2, \eta^{}_3)$ and $[\rho, \sigma]$, the lower bounds of $M^{}_2$ as functions of $m^{}_1$ in the NO case, imposed by requiring the {\bf Scenario I} leptogenesis to be successful. These results are same for $\delta= \pi/2$ and $-\pi/2$.}
\label{fig4.1}
\end{figure*}

\begin{figure*}[t]
\centering
\includegraphics[width=6.7in]{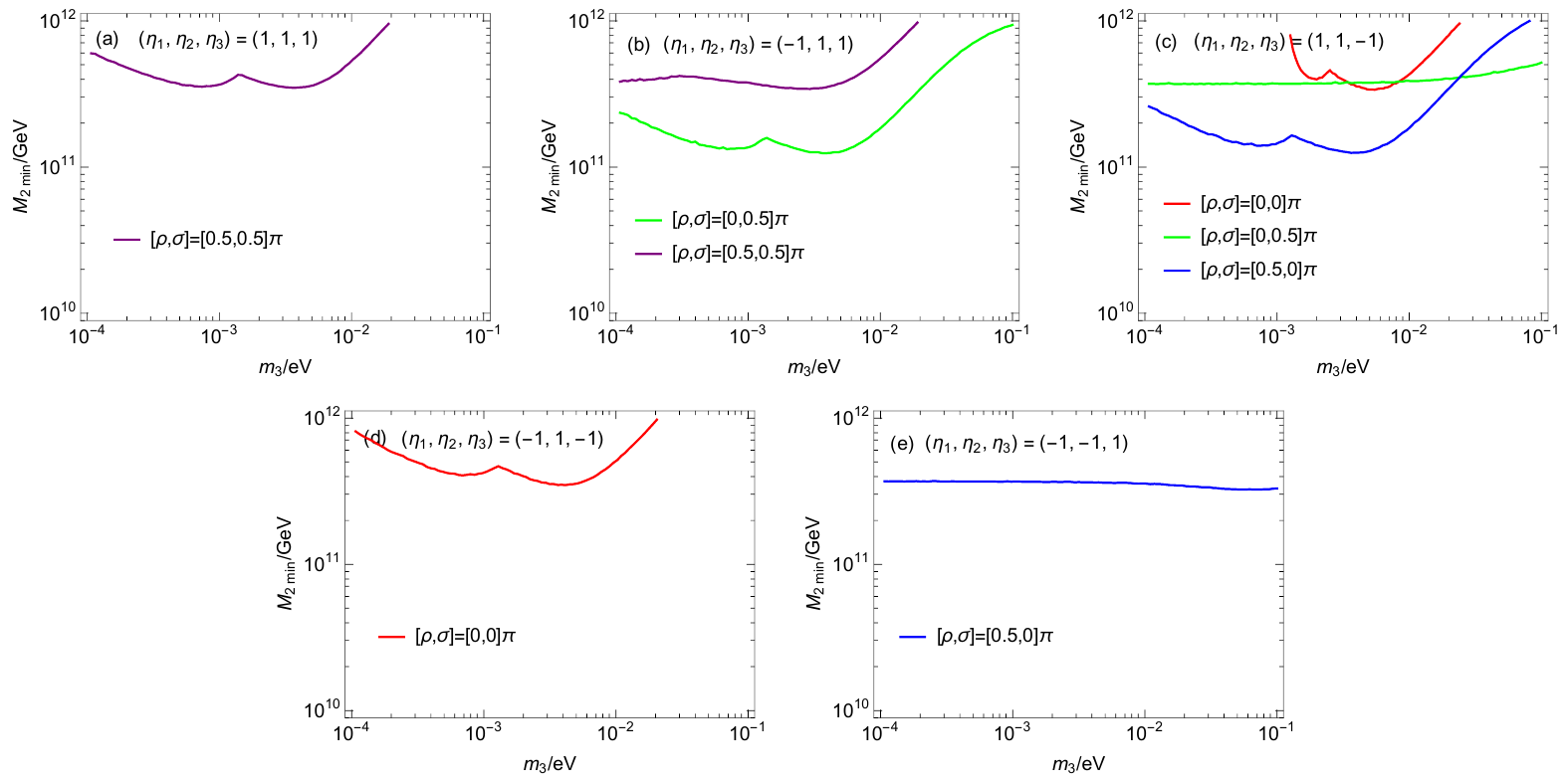}
\caption{ Same as Fig.~\ref{fig4.1}, except these results are for the IO case. }
\label{fig4.2}
\end{figure*}

Depending on the relative sizes of $M^{}_{2, 3}$ and $10^{12}$ GeV, there are the following two possible leptogenesis scenarios. {\bf Scenario I}: for $M^{}_3 > 10^{12}$ GeV $>M^{}_2$, the baryon asymmetry produced from the decays of $N^{}_3$ is vanishing, while that produced from the decays of $N^{}_2$ can be calculated according to Eq.~(\ref{4.10}) with $\varepsilon^{}_{2 \mu}$ as in Eq.~(\ref{4.9}) and $\widetilde m^{}_{2 \mu} = |b^{}_2|^2$. In Fig.~\ref{fig4.1}, for the various combinations of $(\eta^{}_1, \eta^{}_2, \eta^{}_3)$ and $[\rho, \sigma]$ that can accommodate a successful leptogenesis, we plot the lower bounds of $M^{}_2$ as functions of $m^{}_1$ in the NO case. These results are obtained by allowing the unconstrained one of $x$ and $y$ to vary in the whole range.
Note that these results are same for $\delta= \pi/2$ and $-\pi/2$. We see that the requirement of leptogenesis being successful can help us exclude many possible combinations of $(\eta^{}_1, \eta^{}_2, \eta^{}_3)$ and $[\rho, \sigma]$. For the viable combinations of them, in some cases $M^{}_2$ needs to be larger than $10^{11}$ GeV or even close to $10^{12}$ GeV in order for leptogenesis to be successful, while in some other cases leptogenesis has chance to be successful for $M^{}_2 \sim 10^{10}$ GeV and certain values of $m^{}_1$. In Fig.~\ref{fig4.2}, we give the results for the IO case. As in the NO case, only for certain combinations of $(\eta^{}_1, \eta^{}_2, \eta^{}_3)$ and $[\rho, \sigma]$ can leptogenesis have chance to be successful. For these cases, $M^{}_2$ needs to be larger than $10^{11}$ GeV in order for leptogenesis to be successful.

{\bf Scenario II}: for $ 10^{12}$ GeV $>M^{}_3 >M^{}_2$, after the $N^{}_3$-leptogenesis phase, the baryon asymmetry $Y^{0}_{3 \rm B}$ produced from the decays of $N^{}_3$ can also be calculated according to Eq.~(\ref{4.10}) with $\varepsilon^{}_{3 \mu}$ as in Eq.~(\ref{4.9}) and $\widetilde m^{}_{3 \mu} = |c^{}_2|^2$.
After the $N^{}_2$-leptogenesis phase, the surviving amount of $Y^{0}_{3 \rm B}$ can be calculated according to Eq.~(\ref{3.11}) with $\widetilde m^{}_{2 \tau} = |b^{}_2|^2$.
Taking account of the baryon asymmetry $Y^{}_{2 \rm B}$ produced from the decays of $N^{}_2$, which can be calculated can be calculated in a same way as in the former scenario,
the final baryon asymmetry is simply given by $Y^{}_{ \rm B} =Y^{}_{2 \rm B} + Y^{}_{3 \rm B}$. Given that the contribution of $N^{}_3$ to leptogenesis suffers the washout effects of $N^{}_2$,
the numerical results for this scenario are similar to those for the former one, so we will not explicitly show them.

\section{Summary}

In summary, in this paper we have performed a further study of an interesting possibility proposed by Asaka, Ishida and Tanaka that, in spite of the Majorana nature of neutrinos, the $0\nu \beta\beta$ decay can be hidden. In the original AIT model, the AIT ansatz is realized in the minimal seesaw model with two right-handed neutrinos which have a hierarchical mass structure: the lighter and heavier right-handed neutrinos are respectively much lighter and heavier than $\Lambda^{}_{\beta}$. However, the original AIT model does not accommodate a successful leptogenesis. For this problem, in this paper we study a split seesaw model with one lighter right-handed neutrino but two heavier right-handed neutrinos which can realize the AIT ansatz and accommodate a successful leptogenesis simultaneously.

We have first given the condition on the neutrino Yukawa couplings [i.e., $(Y^{}_\nu)^{}_{e2} = (Y^{}_\nu)^{}_{e3} =0$] for realizing the AIT ansatz [see Eq.~(\ref{2.1})], discussed its realization by employing an Abelian flavor symmetry, and studied its implications for the mixing strengths $|\Theta^{}_{\alpha 1}|$ of the lighter right-handed neutrino with three left-handed neutrinos. It is interesting that $|\Theta^{}_{\alpha 1}|$ can be completely determined from the low-energy neutrino observables (plus $M^{}_1$). We have then successively studied the implications for leptogenesis of the interesting scenarios where $M^{}_{\rm D}$ is a triangular matrix (which has maximally-restricted texture zeros, in line with the simplicity principle) or respects the $\mu$-$\tau$ reflection symmetry (which is well motivated by the experimental results), on top of the AIT ansatz.

In the scenario of $M^{}_{\rm D}$ being a triangular matrix, the model parameters can be completely reconstructed from the low-energy neutrino observables [see Eq.~(\ref{2.9})]. In this case, the low-energy neutrino observables will be subject to the constraints from leptogenesis.
For the purpose of illustration, we have only studied the cases where only one of $\delta$, $\rho$ and $\sigma$ acts as the source for CP violation while the other two of them simply take trivial values. For all the possible cases we have given the lower bounds of $M^{}_2$ as functions of the lightest neutrino mass that are obtained by requiring leptogenesis to be successful. It is found that leptogenesis is more difficult to be successful when $\delta$ is the only source for CP violation. In particular, in the IO case, when $\delta$ is the only source for CP violation, the ranges of $M^{}_2$ for leptogenesis to be successful are very small. This can be easily understood from that the effects of $\delta$ are always suppressed by $s^{}_{13}$. This means that in this case the requirement of leptogenesis being successful will help us falsify the possibility that $\delta$ is the only source for CP violation. Furthermore, to see more clearly the constraints on the CP phases from leptogenesis, for some benchmark values of $M^{}_2$ we have plotted the values of the CP phases versus the lightest neutrino mass for leptogenesis to be successful.

In the scenario of $M^{}_{\rm D}$ respecting the $\mu$-$\tau$ reflection symmetry [see Eq.~(\ref{4.5})] on top of the AIT ansatz, the lepton flavor mixing parameters are predicted to take some special values as given in Eq.~(\ref{4.3}), and the model parameters can be reconstructed from the low-energy neutrino observables in combination with the unconstrained one of $x$ and $y$ [see Eqs.~(\ref{4.7}, \ref{4.8})]. Because of the $\mu$-$\tau$ reflection symmetry, only in the two-flavor regime can leptogenesis have chance to be successful. It is found that leptogenesis has no chance to be successful for some combinations of $(\eta^{}_1, \eta^{}_2, \eta^{}_3)$ and $[\rho, \sigma]$. This means that the requirement of leptogenesis being successful will help us falsify such parameter options.
For the viable combinations of $(\eta^{}_1, \eta^{}_2, \eta^{}_3)$ and $[\rho, \sigma]$, we have given the lower bounds of $M^{}_2$ as functions of the lightest neutrino mass that are obtained by requiring leptogenesis to be successful.

Our results show that a split seesaw model with one lighter right-handed neutrino and two heavier right-handed neutrinos can realize the AIT ansatz, accommodate a successful leptogenesis and give viable  low-energy neutrino observables simultaneously.

Before ending the paper, we would like to give three comments as follows. (1) The particular scenario [i.e., $M^{}_1 \ll \Lambda^{}_{\beta} \ll M^{}_2 < M^{}_3$ and $(M^{}_{\rm D})^{}_{e2} = (M^{}_{\rm D})^{}_{e3} =0$] studied in the present work is different from the well-known scenario that $m^\nu_{\beta \beta}$ [i.e., $(M^{}_\nu)_{ee}$] itself can vanish: on the one hand, in the former scenario the vanishing of $m^{}_{\beta \beta}$ results from the cancellation between $m^\nu_{\beta \beta}$ and $m^N_{\beta \beta}$, while in the latter scenario it results from the vanishings of both $m^\nu_{\beta \beta}$ and $m^N_{\beta \beta}$; on the other hand, in the former scenario $m^{}_{\beta \beta}$ vanishes independent of the concrete values of the neutrino mass parameters, while in the latter scenario $m^\nu_{\beta \beta}$ can only vanish for certain values of the neutrino parameters in the case of the neutrino masses being of the normal ordering \cite{Vissani}. (2) The leptogenesis in the present framework is different from that in the minimal seesaw model (where there are also only two right-handed neutrinos relevant for leptogenesis) \cite{MSS, MSS2}: in the minimal seesaw model, only two right-handed neutrinos contribute to the mass generation of light neutrinos, rendering one light neutrino to keep massless; in the present framework, all the three right-handed neutrinos do so [see the discussion below Eq.~(\ref{2.5})], allowing all the three light neutrinos to be massive; consequently, the neutrino parameter space explored in the present work is different from that in the minimal seesaw model. (3) In the mass region of interest, the lifetime of $N^{}_1$ is so long that its decays would destroy the success of the big-bang nucleosynthesis. As pointed out by Ref.~\cite{hiding}, one possibility to avoid this issue is the dilution of the $N^{}_1$ abundance by the late-time entropy production. Note that the two heavier right-handed neutrinos (whose masses are not nearly degenerate) cannot take the burden for the late-time entropy production, since they need to be above the Davidson-Ibarra bound in order to accommodate a successful leptogenesis so that they decay at a temperature far above the electroweak scale. There are two possibilities to overcome this issue. One possibility is that the late-time entropy production arises from some particles which are dedicated to dilute the unwanted relics like the gravitino, the Polonyi, the moduli  and the dilaton figure in supersymmetric and string theory models \cite{GR}. The other possibility is that the two heavier right-handed neutrinos are also lowered to the electroweak scale so that they can be responsible for the late-time entropy production as in Ref.~\cite{hiding}. But it should be noted that in this scenario their masses need to be nearly degenerate in order to realize the resonant leptogenesis scenario \cite{resonant}.

\vspace{0.5cm}

\underline{Acknowledgments} \vspace{0.2cm}

This work is supported in part by the National Natural Science Foundation of China under grant Nos.~11605081, 12142507 and 12147214, and the Natural Science Foundation of the Liaoning Scientific Committee under grant NO.~2022-MS-314.

\end{document}